\documentclass[prd,aps,floats,floatfix]{revtex4}
\usepackage{amsmath}
\usepackage{amsfonts}
\usepackage{graphicx}
\usepackage{epsfig}
\usepackage{epstopdf}
\begin{document}

\preprint{MSUHEP-080606}
\title{Lone Higgs boson at the CERN LHC}
\author{Ken Hsieh}
\email[]{kenhsieh@pa.msu.edu}
\author{C.--P. Yuan}
\email[]{yuan@pa.msu.edu}
\affiliation{Department of Physics and Astronomy, Michigan State
University, East Lansing, MI 48824, USA}
\date{\today}
\begin{abstract}
We address the possible scenario that the Large Hadron Collider
(LHC) discovers only a Higgs boson after 10 fb$^{-1}$ of
operation, and attempt to identify this Higgs boson as that of the
Standard Model (SM), the minimal universal extra dimension model
(MUED), the littlest Higgs model with $T$-parity (LHT), or the
minimal supersymmetric Standard Model (MSSM), using only the
measurement
of the product of gluon-fusion production cross section
and the di-photon branching ratio.
In MUED, by decoupling any new physics sufficiently
to evade the discovery reach at the LHC,
the deviation of the signal from the SM is not
statistically significant.
However, in LHT and MSSM, it is possible to
have a significant deviation in the signal that
is consistent with this "lone Higgs scenario",
and, in the case of a very large suppression,
we can distinguish MSSM and LHT before the
discovery of any
new resonances.
Starting with the lone Higgs scenario and the deviation
in this measurement from the Standard Model prediction
(whether or not statistically significant),
we offer tests that may discriminate the models and
search strategies of discovering
new physics signatures
with increasing integrated luminosity.
\end{abstract}

\newcommand{\HGG}{B\sigma(gg\rightarrow h\rightarrow\gamma\gamma)}
\newcommand{\vect}[1]{\overrightarrow{#1}}
\newcommand{\smbox}[1]{\mbox{\scriptsize #1}}
\newcommand{\tbox}[1]{\mbox{\tiny #1}}
\newcommand{\vev}[1]{\langle #1 \rangle}
\newcommand{\Hbar}{\overline{H}}
\newcommand{\Tr}[1]{\mbox{Tr}\left[#1\right]}
\newcommand{\invfb}{\mbox{fb}^{-1}}

\maketitle
\newpage

\section{Introduction}
The stability of the electroweak scale has driven the high energy physics
community, both theorists and experimentalists alike,
for nearly the past two decades.
With the advent of the Large Hadron Collider (LHC), we
can finally probe the mechanism of electroweak symmetry breaking
(EWSB) and possibly new physics at the TeV scale that stabilizes
the electroweak scale.
However, as such new physics is still a mystery, we
need to be prepared for all the possibilities.
In addition, with
the multitude of models of new physics and the possible associated
experimental signatures, we are also faced with the `inverse problem' of
distinguishing models of new physics using the experimental data.

In this work, we investigate one of the possible scenarios at the
LHC, and attempt to disentangle three generic models of new
physics based on experimental measurements of a Higgs boson.
We suppose that, after the first few years of operation with an integrated
luminosity of 10 $\invfb$, the LHC has only discovered a lone
scalar boson with couplings to the $W$- and $Z$-bosons that are of
the same magnitude as predicted in the Standard Model (SM).
While discovering only a Higgs boson at the LHC (with an integrated
luminosity of several 100 $\invfb$)
has been dubbed a ``Nightmare Scenario \cite{Nightmare},''
here we are only assuming no new physics, other than this
Higgs boson, is seen at this stage of operation of the LHC,
and leave open the possibility that new physics may be
uncovered with further operation time.
Indeed, one of the main goals of works of this type is to optimize
further search strategies based on the information we have at hand from
the discovered Higgs boson.

Let us denote $\HGG$ as the product of the Higgs boson production
cross section $\sigma(pp\rightarrow(gg\rightarrow h)X)$ and the
di-photon decay branching ratio
$\mbox{Br}(h\rightarrow\gamma\gamma)$.
The main question that we attempt to answer in this work is: from
the measurement of $\HGG$ and its deviation from the SM
prediction, can we identify this scalar boson as \emph{the} Higgs
boson in the SM, minimal universal extra dimensions (MUED),
littlest Higgs with $T$-parity (LHT), or the lightest $CP$-even
boson in the minimal supersymmetric standard model (MSSM)?
If not, we investigate whether we can use this measurement
as a hint or bias, and devise further search strategies
of new physics based on its deviation from the SM, regardless
whether such deviation is statistically significant.
Questions of this type are in spirit similar to the LHC
inverse problem \cite{ArkaniHamed:2005px},
but with an emphasis on distinguishing
the models rather than mapping the regions of parameter
spaces of a particular model from the data.
While such LHC inverse problems have been studied
in the literature, they attempt to distinguish
models through properties, such as spin,
of the new resonances discovered at the LHC.
Our work here is also of similar spirit to Mantry et al.~\cite{Mantry:2007sj}
\cite{Mantry:2007ar} and Randall~\cite{Randall:2007as}, where
they discuss how the properties of the Higgs boson can be modified due
to states that are not directly observable at the LHC.

Our work is organized as follows.
In Section \ref{sec:precision}, we discuss the precision
to which the signal $\HGG$ can
be measured at the LHC after 10 $\invfb$
of data.
In Section \ref{sec:HGG-model}, we discuss
the general pattern of deviations of the signal $\HGG$
in the parameter spaces of the models, and roughly
map out regions in parameter spaces that such
deviation can be significant.
We also apply the results of LHC reaches in these models
to map out regions of parameter spaces that can
be consistent with the aforementioned lone Higgs scenario.
In Section \ref{sec:HGG-lone},
we apply the lone Higgs scenario as constraints on
the parameter spaces, and see how the signal is affected.
In particular, we find that in the lone Higgs scenario with a
large deviation in $\HGG$,
we can potentially rule out MUED, and,
in some cases, distinguish between the MSSM and LHT.
Also in Section \ref{sec:HGG-lone}, we propose some parameter-independent
tests
that can also be used to distinguish these models.
We conclude in Section \ref{sec:conc} with a
summary of our results and offer outlook
for projects of this type.

\section{A Review of Higgs Measurements at the LHC}
\label{sec:precision}
In this section we present a brief overview of the
detection of the Higgs boson and the measurement
of its properties at the LHC.  More details
can be found at the ATLAS technical design report (TDR)
\cite{:1999fr} and CMS TDR \cite{Ball:2007zza}, and references therein.

\begin{figure}[h!t]
\begin{center}
\includegraphics [width=4.0in]{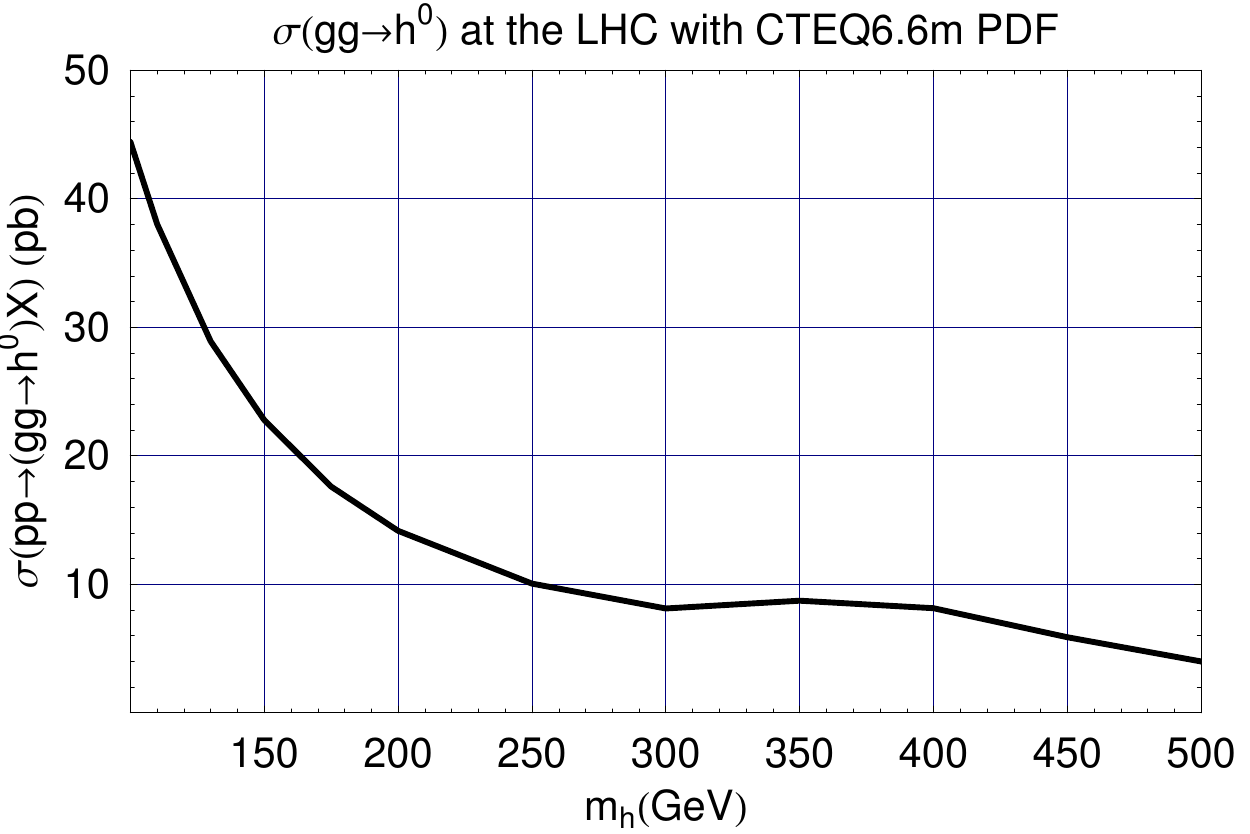}
\includegraphics [width=4.0in]{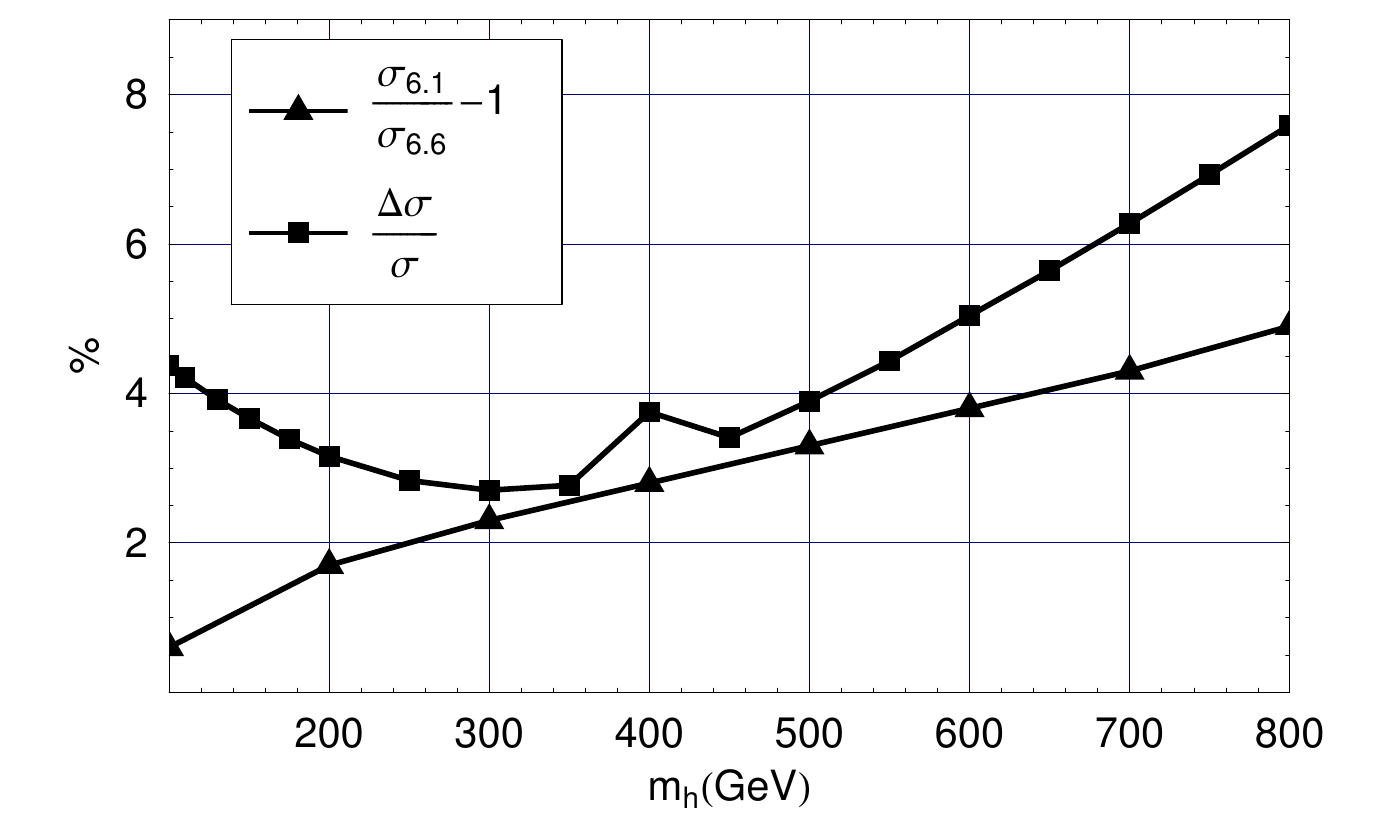}
\caption{
The cross section $\sigma_h$
at the LHC using the latest PDF, CTEQ6.6M \cite{Nadolsky:2008zw}
(top plot) and its PDF-induced uncertainty (bottom plot, boxed points)
and relative difference with previous version of PDF, CTEQ6.1
(triangle points).
}
\label{fig:CTEQ66GGH}
\end{center}
\end{figure}

For reference, we show the production cross section
of the Higgs boson via gluon-gluon fusion at the LHC
$\sigma_h\equiv\sigma(pp\rightarrow(gg\rightarrow h^0)X)$
at the next-to-leading order in QCD, with
the renormalization and factorization scales set to the mass
of the Higgs boson ($m_h$),
using the latest parton distribution functions
(PDF), CTEQ 6.6M \cite{Nadolsky:2008zw}, in the top plot
of Fig.~\ref{fig:CTEQ66GGH}.
The uncertainties of this cross section,
both the PDF-induced uncertainty
as well as the relative difference with an
earlier version of the PDF (CTEQ6.1), are of
the order of a few percent as shown in the lower plot
of Fig.~\ref{fig:CTEQ66GGH}.
The uncertainty in the luminosity will
be on the order of 20\% at the start of the LHC.
However, the uncertainties in the measurements of the
cross sections due to the
uncertainty in the luminosity can be reduced
partially by taking ratios of these cross sections
to measured ``standard candle'' cross sections, such as
$\sigma_{t\overline{t}}\equiv\sigma(pp\rightarrow t\overline{t}X)$ and
$\sigma_Z\equiv\sigma(pp\rightarrow
(Z^0\rightarrow \ell^{+}\ell^{-})X)$ \cite{Nadolsky:2008zw}.

The detection channels of the Higgs boson depend significantly on
its mass.
Although the Higgs boson couples most strongly
to the massive gauge bosons $W^{\pm}, Z^{0}$ and the top
quark, for Higgs mass significantly
lighter than the $WW$ threshold ($m_h \lesssim 130$ GeV),
the decays $h\rightarrow WW, ZZ, \overline{t}t$ are kinematically
inaccessible, and
the dominant decay channel of the Higgs
boson is $h\rightarrow\overline{b}b$.
Unfortunately, the di-jet background at the LHC is expected
to overwhelm this signal, and the most promising channel of
detecting the Higgs boson is through its (loop suppressed) di-photon
decay, $h\rightarrow\gamma\gamma$, with a branching ratio of
about 0.2\%.
The di-photon channel offers a very clean
signature of Higgs boson and enables a precise
measurement of its mass.
For $m_h > 130$ GeV, the decay channels $W^{(\ast)}W$ and
$Z^{(\ast)}Z$, and (for even heavier $m_h$) $\overline{t}t$
become dominant.
At the same time, the di-photon branching ratio decreases
significantly.

As we are assuming that the discovered Higgs boson
is (maybe only one of several Higgs bosons)
responsible for electroweak symmetry breaking, the
most general renormalizable operators
involving this Higgs boson would be those in
the SM.
The loop contributions to these
couplings from new physics will typically
be of the percent level that are too small to be probed
at the early stages of the LHC, and we may ignore the loop
corrections to these couplings.
On the other hand, the couplings $a_{hgg}$ and $a_{h\gamma\gamma}$ of dimension-five di-gluon and
di-photon operators that
characterize the gluon-fusion production rate
and di-photon width of the Higgs boson
\begin{align}
a_{hgg} \frac{h}{v_{\smbox{ew}}} G^{A}_{\mu\nu}G^{A\mu\nu}, \quad
a_{h\gamma\gamma}\frac{h}{v_{\smbox{ew}}} F_{\mu\nu}F^{\mu\nu},
\end{align}
are one-loop at leading order in the SM, and loop contributions
from new physics may be competitive.
Thus, to study this discovered boson in a bottom-up approach,
we consider an effective Lagrangian that includes
all the renormalizable gauge and Yukawa operators as in the SM,
but with arbitrary coefficients, and only consider the leading-order
effects of these operators.
In addition, we include
the two dimension-five operators with arbitrary coefficients
that parameterizes the leading effects of the yet-undiscovered
new physics on the Higgs boson, and the measurement of $\HGG$ essentially
measures the product of $a^2_{hgg}$ (which is proportional
to the production cross section $\sigma_h$) and the decay branching ratio
$\mbox{Br}(h\rightarrow\gamma\gamma)$.
As the di-photon branching ratio is only significant
for $m_h\lesssim 130$ GeV, we will only
consider a Higgs boson with a mass within this range.

Since the PDF-induced uncertainty is of the order of 5\%,
the precision to which these couplings can be measured
depends crucially on the uncertainty in the luminosity
at the LHC.
The precision of which the couplings in our effective Lagrangian
can be measured at the LHC has been extensively studied
\cite{Zeppenfeld:2000td}
\cite{Duhrssen:2004cv}.
From Zeppenfeld et al.~\cite{Zeppenfeld:2000td}, we see that
with 100 $\invfb$ from both ATLAS \cite{:1999fr} and CMS \cite{Ball:2007zza},
the cross section $\HGG$ can be measured to about 10\%.
This uncertainty is defined as
\begin{align}
\frac{\sqrt{N_S+N_B}}{N_S},
\end{align}
where $N_S(N_B)$ is the number of signal (background) events.
We refer the readers to the reference for the numbers
of events, backgrounds, and the significance of
the signal.

With only 10 $\invfb$ of data, we naively scale our error
by a factor of $\sqrt{0.1}/0.1\sim 3$, and use 30\% as
the accuracy to which $\HGG$ can be measured.
Thus, the measurement of $\HGG$ with
10 $\invfb$ at LHC can only
distinguish  models of new physics
from the SM only if it deviates by
more than 30\% from the SM prediction,
and we will see that this can
often places stringent
constraints on the parameter
spaces of new physics models,
independent of the lone Higgs
scenario.

\section{Models of New Physics and Lone Higgs Scenarios}
\label{sec:HGG-model}
\subsection{Minimal Supersymmetric Standard Model}
\subsubsection{$\HGG$ in MSSM}
\label{sec:MSSM-HGG}
The minimal supersymmetric Standard Model (MSSM) is the most
widely-studied model of new physics both
in theory and experiment~\cite{Martin:1997ns}.
It extends the
SM with superpartners that differ in spin by 1/2 from their SM counterparts,
and the electroweak scale is stabilized by the presence of these superpartners
if they have masses of about 1 TeV.
It is remarkable that in the MSSM the lightest $CP$-even boson
has a mass that is bounded by about 125 GeV if the MSSM is to solve the
hierarchy problem.
As decay to $WW^{\ast}$ and $ZZ^{\ast}$ pairs are now kinematically suppressed,
the di-photon channel is now the golden channel to search for the
lightest $CP$-even Higgs boson.

\begin{figure}[h!t]
\begin{center}
\includegraphics [width=2.7in]{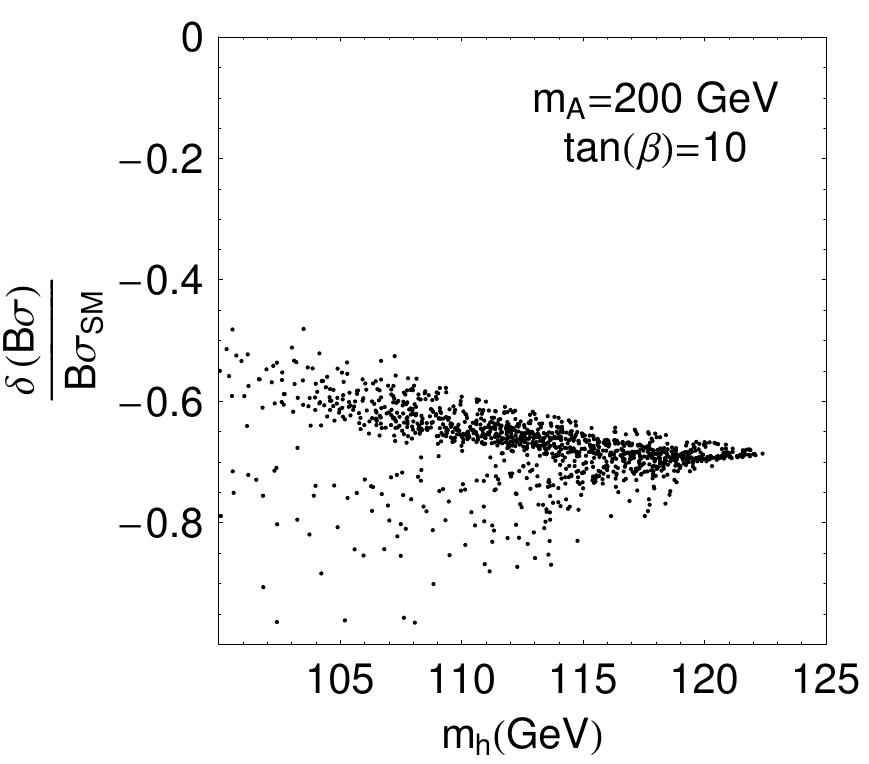}
\includegraphics [width=2.7in]{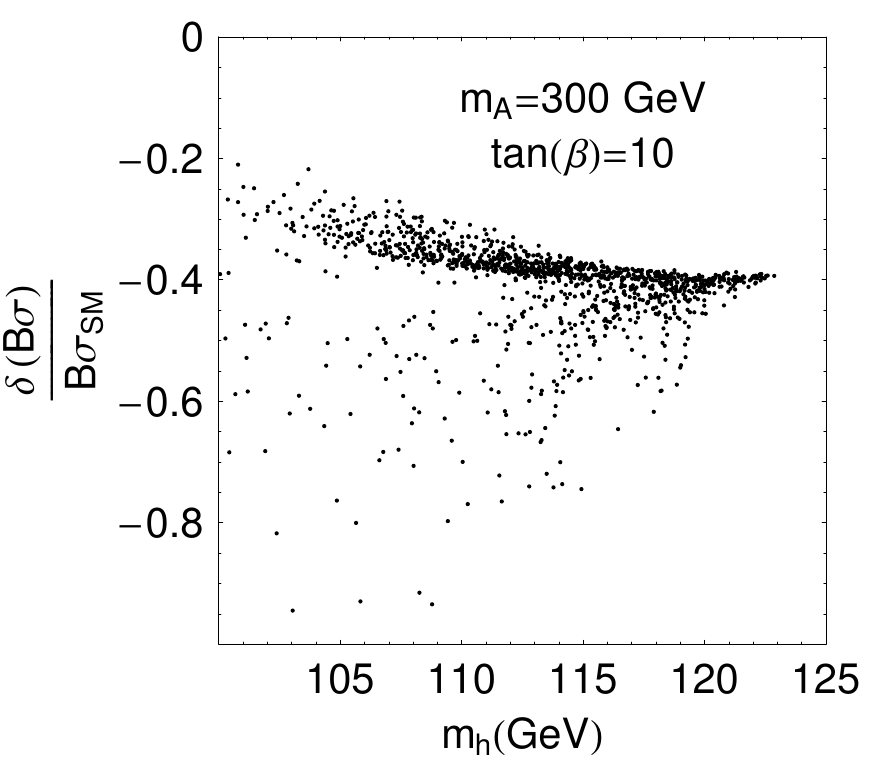}\\
\includegraphics [width=2.7in]{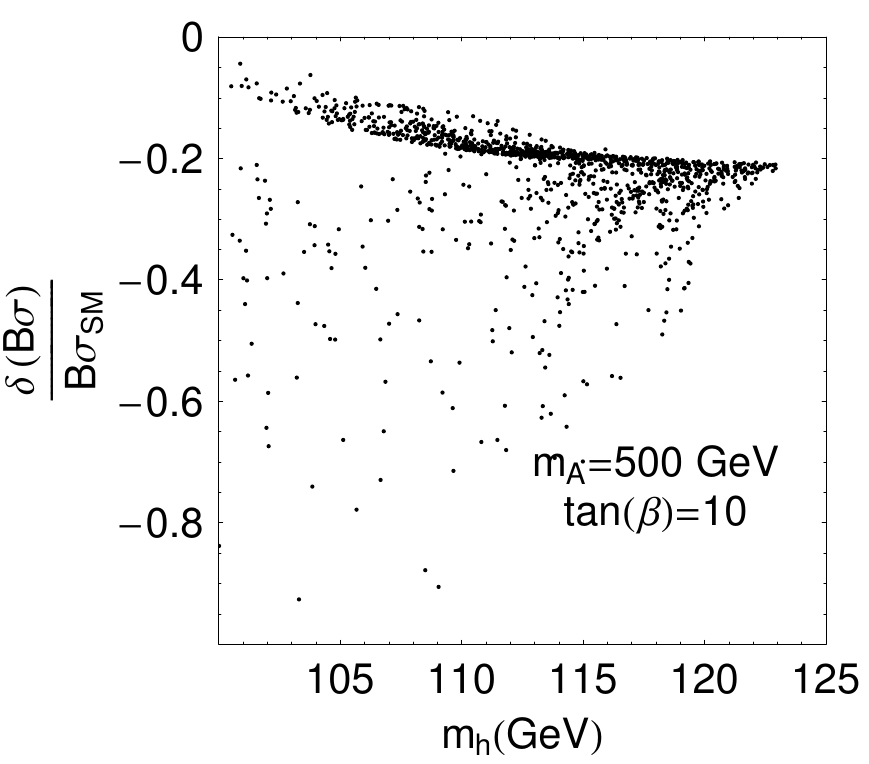}
\includegraphics [width=2.7in]{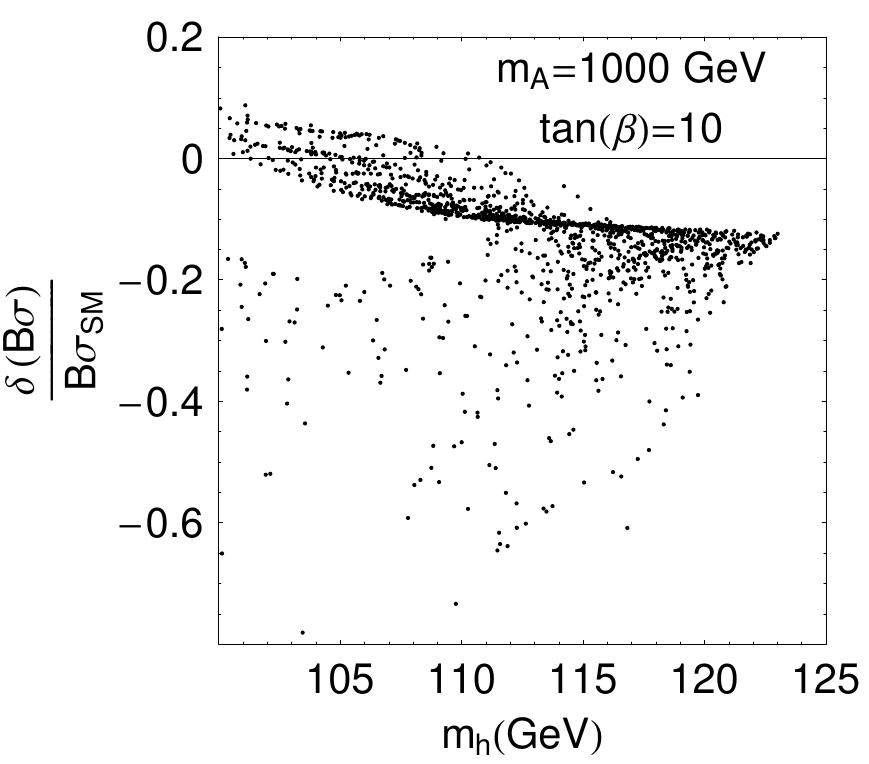}
\caption{The fractional deviation of $\HGG$ as a function
of Higgs mass in MSSM for various values of $M_A$=200 (top left),
300 (top right), 500 (bottom left), and  1000 (bottom right) GeV.
The s-top soft masses $M_{\tilde{Q}_3}$ and
$M_{\tilde{\overline{U}}_3}$
are scanned from 300 GeV to 1.5 TeV, with $A_t$ scanned
in the range of $\pm 4(M_{\tilde{Q}_3}M_{\tilde{\overline{U}}_3})^{1/2}$.
The other SUSY-breaking values are fixed as $M_{\tilde{\ell}}=100$ GeV,
$M_{\tilde{w}}=\mu=200$ GeV, and $M_{\tilde{g}}=M_{\tilde{Q}}=500$ GeV.
}
\label{fig:MSSMPlot}
\end{center}
\end{figure}

\begin{figure}[h!bt]
\begin{center}
\includegraphics [width=3.5in]{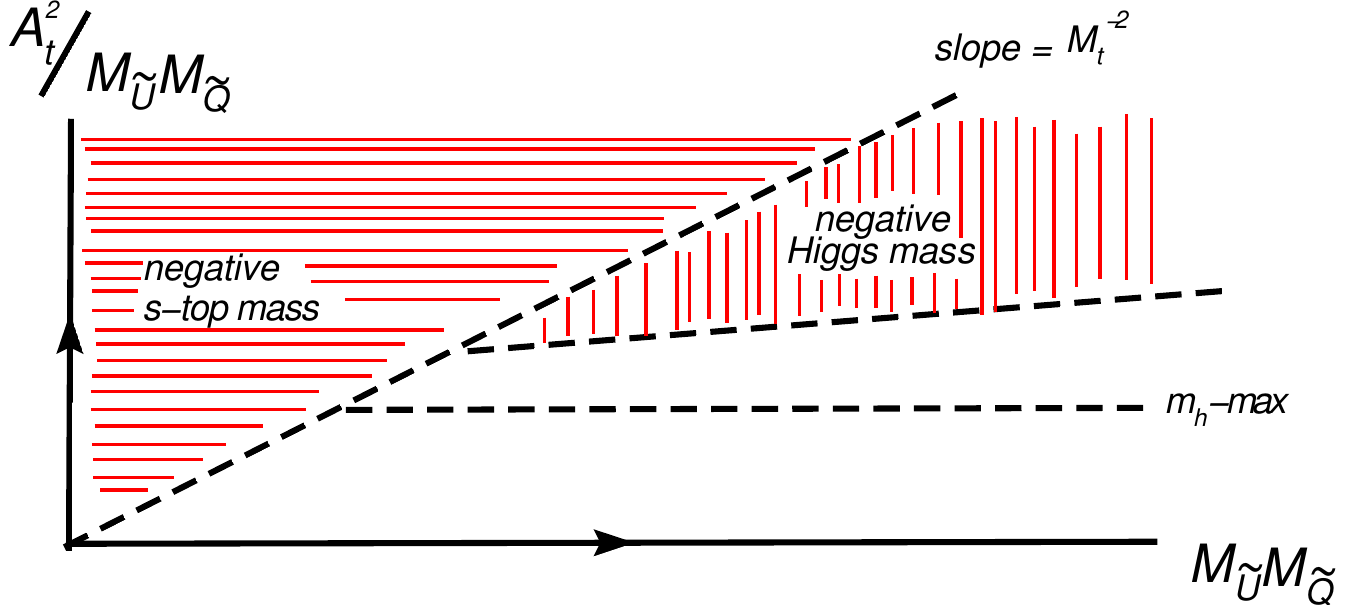}
\caption{The schematic plot of allowed parameter space in the s-top sector.
This plot is not drawn to scale.
The line denoted as $m_h\!\!-\!\!\mbox{max}$ denotes the line
with $A^2_t/M_{\tilde{Q}_3}M_{\tilde{\overline{U}}_3}=6$,
and the mass of the lightest, $CP$-even Higgs boson is maximized
at 1-loop.
Having too large $A_t$ can lead to a negative s-top mass
(horizontally hashed region).
However, even if $A_t$ is not large enough to lead
to a negative s-top mass, it can be large enough
to
lead to a negative mass for the lightest, $CP$-even Higgs boson
(vertically hashed region).
}
\label{fig:s-top-param}
\end{center}
\end{figure}

Unfortunately, the MSSM comes with 105 parameters \cite{Martin:1997ns} and it is impossible
to scan through such vast parameter space.  We will make
some simplifying assumptions that the first two generations of
sfermions have mass matrices that are diagonal at the weak scale
and the phases of all SUSY-breaking contributions are zero.
These assumptions are consistent with the various flavor-changing
experimental constraints, and leave us with a reduced parameter space.
With this reduced parameter space,
the signal $\HGG$ can still vary greatly.
For example, in Fig.~\ref{fig:MSSMPlot}, we use
\verb"hdecay" \cite{Djouadi:1997yw} (which includes
the \verb"FeynHiggs" package \cite{Heinemeyer:1998yj})
to show the deviation of $\HGG$ from the SM
values for various values of $M_A$ (the mass of
the $CP$-odd Higgs boson in the MSSM),
scanning over the s-top sector
parameters
\begin{eqnarray}
300\ \mbox{GeV} \leq &M_{\tilde{Q}_3}&
\leq 1.5\ \mbox{TeV},\nonumber\\
300\ \mbox{GeV} \leq &M_{\tilde{\overline{U}}_3}&
\leq 1.5\ \mbox{TeV},\nonumber\\
-4\sqrt{M_{\tilde{Q}_3}M_{\tilde{\overline{U}}_3}} \leq &A_t &\leq
4\sqrt{M_{\tilde{Q}_3}M_{\tilde{\overline{U}}_3}},\nonumber
\end{eqnarray}
where $M_{\tilde{Q}_3}$ and $M_{\tilde{\overline{U}}_3}$
are respectively the SUSY-breaking masses of the
left- and right-handed s-tops (the superpartners
the top quark),
and $y_tA_t$ (where $y_t$ is the top Yukawa coupling) is the coefficient of the trilinear
interaction $\tilde{Q}_3H_u^0\tilde{\overline{U}}_3$.
We note the following points regarding the parameters in
the s-top sector (also see Fig.~\ref{fig:s-top-param}).
\begin{itemize}
\item
The scanned range of $A_t$ includes the regions
that give the largest mass for the lightest, $CP$-even Higgs boson,
which occurs for $A^2_{t}\sim 6M_{\tilde{Q}_3}M_{\tilde{\overline{U}}_3}$.
While
$A^2_{t}\sim 6M_{\tilde{Q}_3}M_{\tilde{\overline{U}}_3}$
leads a large Higgs mass, the mixing in the s-top
sector is $m_t (A_{t}-\mu \cot\beta)$,
so we can have a large Higgs mass without
having a light s-top with
$M_{\tilde{Q}_3},M_{\tilde{\overline{U}}_3}\gg m_t$.
\item
To avoid s-tops with negative squared-mass, $A_t$ must satisfy (for large $\tan\beta$)
\begin{align}
A_t^2 < \frac{(M_{\tilde{Q}_3}^2+m_t^2)(M^2_{\tilde{\overline{U}}_3}+m_t^2)}{m_t^2}.
\end{align}
For small $M_{\tilde{Q}_3}$ and $M_{\tilde{\overline{U}}_3}$,
it may not be
possible to scan in the full region between
$A_t\sim \pm 4(M_{\tilde{Q}_3}M_{\tilde{\overline{U}}_3})^{1/2}$.
For small $M_{\tilde{Q}_3}$ and $M_{\tilde{\overline{U}}_3}$,
our scanned range in $A_t$ is limited requiring having two
s-tops with positive masses.
\item
With large $M_{\tilde{Q}_3}$ and $M_{\tilde{\overline{U}}_3}$, even
though positive s-top masses may allow $A^2_t$ to be large relative to
$M_{\tilde{Q}_3}M_{\tilde{\overline{U}}_3}$,
large $A_t$ can lead to a negative squared mass for the lightest,
$CP$-even Higgs boson.
This occurs when we have
\begin{align}
\frac{3y_t^4v^2_{\smbox{ew}}}{2\pi^2}\left(
\frac{1}{12}\frac{A_t^4}{M^2_{\tilde{Q}_3}M^2_{\tilde{\overline{U}}_3}}
-\frac{A_t^2}{M_{\tilde{Q}_3}M_{\tilde{\overline{U}}_3}}\right)
\gtrsim
M_Z^2+\frac{3y_t^4v^2_{\smbox{ew}}}{2\pi^2}
\ln\frac{M_{\tilde{Q}_3}M_{\tilde{\overline{U}}_3}}{m_t^2},
\end{align}
where $v_{\smbox{ew}}(=246\ \mbox{GeV})$ is the electroweak scale
(i.e.~the vacuum expectation value),
and our scanned range of $A_t$ does not include such
large $A_t$.
\end{itemize}
For simplicity, we hold all other parameters fixed as
\begin{align}
M_{\tilde{\ell}}=&100\ \mbox{GeV},\nonumber\\
M_{\tilde{w}}=\mu=&200\ \mbox{GeV},\nonumber\\
M_{\tilde{g}}=M_{\tilde{Q}}=&500\ \mbox{GeV},
\label{eq:ScannedSet}
\end{align}
where $M_{\tilde{\ell}}(M_{\tilde{Q}})$
is any slepton (first two generations squark) soft mass, $M_{\tilde{g}}$ the gluino mass,
$M_{\tilde{w}}$ the wino mass, and $\mu$ the chargino mass.
The bino mass
$M_{\tilde{b}}$ is determined assuming unified gaugino
mass at the grand unification scale
$M_{\smbox{GUT}}\sim 10^{16}$ GeV.
%


Though there is a general trend of suppression in $\HGG$
in Fig.~\ref{fig:MSSMPlot}, we see that the specific amount
of suppression fluctuates with the parameters
in the s-top sector
as well as $M_A$.
(We discuss the region where $\HGG$ is enhanced
in the lower-right plot of Fig.~\ref{fig:MSSMPlot}
at the end of this subsection.)
However, as we eventually will be interested in
the lone Higgs scenario, the s-fermions and the
gauginos should be heavy enough
to evade discovery.
In particular, having large s-top masses and
large $A^2_{t}(\sim 6M_{\tilde{Q}_3}M_{\tilde{\overline{U}}_3})$
leads to a
large Higgs mass
in addition to heavy s-top
masses to evade discovery.

\begin{figure}[h!t]
\begin{center}
\includegraphics [width=4.25in]{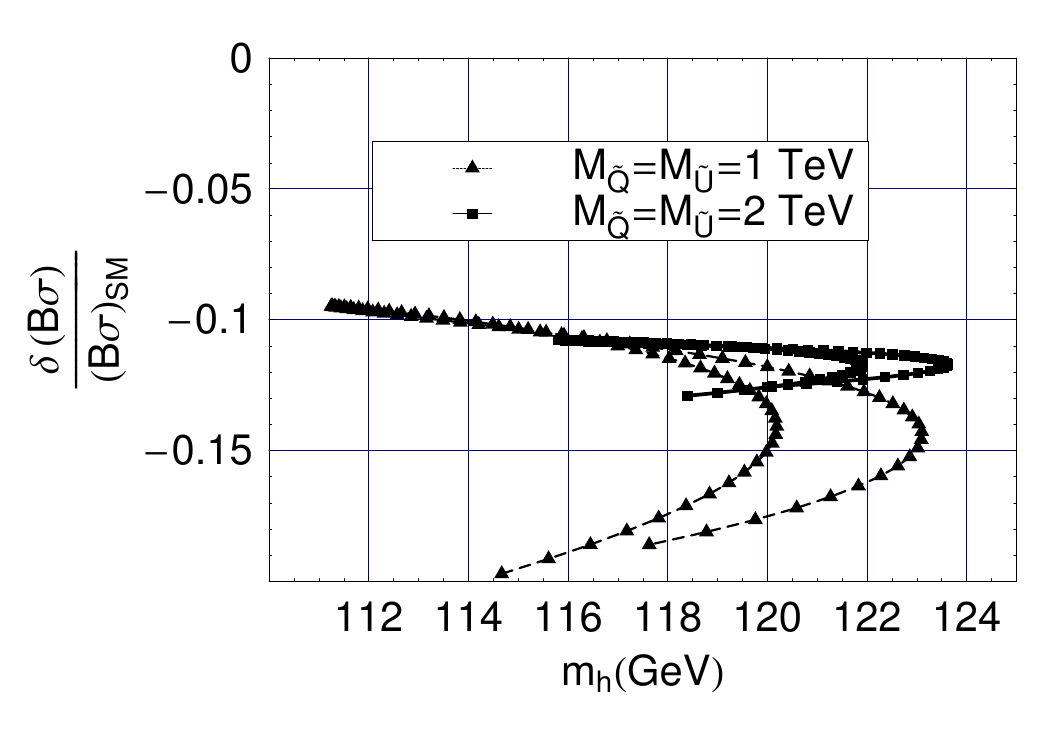}
\caption{The fractional deviation of $\HGG$ with
$A_t$ scanned over the range
$\pm3(M_{\tilde{Q}_3}M_{\tilde{\overline{U}}_3})^{1/2}$
with $M_{\tilde{Q}_3}=M_{\tilde{\overline{U}}_3}=1\ \mbox{TeV}$ (triangle points),
and $M_{\tilde{Q}_3}=M_{\tilde{\overline{U}}_3}=2\ \mbox{TeV}$ (square points).
With larger $M_{\tilde{Q}_3,\tilde{\overline{U}}_3}$, the deviation
in $\HGG$ is less sensitive to the variations in $A_t$ because
the s-top sector decouples.
Here we fix $\tan\beta=10$ and $M_A=1$ TeV in addition
to those labelled in Eq.~(\ref{eq:ScannedSet}).
For each set of s-top sector parameters, there are
two arcs on the plot because positive values
of $A_t$ gives a larger Higgs mass than negative values
of $A_t$ at two-loop.
}
\label{fig:runST}
\end{center}
\end{figure}

In Fig.~\ref{fig:MSSMPlot}, we see
that, as $m_h$ increases,
the fluctuation in $\HGG$ decreases,
signaling the decoupling
of the s-top sector.
We show this more explicitly in Fig.~\ref{fig:runST} where we
scan over
$A_{t}$ in the range of $\pm 3 (M_{\tilde{Q}_3}M_{\tilde{\overline{U}}_3})^{1/2}$
holding $M_{\tilde{Q}_3}$ and $M_{\tilde{\overline{U}}_3}$ fixed.
We also fix other parameters as Eq.~(\ref{eq:ScannedSet}),
and fix $\tan\beta=10$ and $M_A$=1 TeV.
With $M_{\tilde{Q}_3}$ and $M_{\tilde{\overline{U}}_3}$ fixed as
1 TeV (2 TeV), the fractional deviation in $\HGG$ changes by 11\% (3\%)
as we scan over $A_t$.
The plot shows that with larger $M_{\tilde{Q}_3,\tilde{\overline{U}}_3}$,
the deviation in $\HGG$ is less sensitive to the variation in $A_t$
as both the s-tops decouple.
We therefore consider a limited lone Higgs scenario
where the s-top soft masses are large enough that,
in addition to evade direct discovery at the LHC,
the s-top contributions to the gluon-gluon fusion and di-photon decay
amplitudes decouple regardless of the value of $A_t$.
As the s-top contributions decouple, we attribute the suppression in $\HGG$
to $M_A$ and $\tan\beta$ in this limited lone Higgs scenario.

\begin{figure}[h!t]
\begin{center}
\includegraphics [width=4.25in]{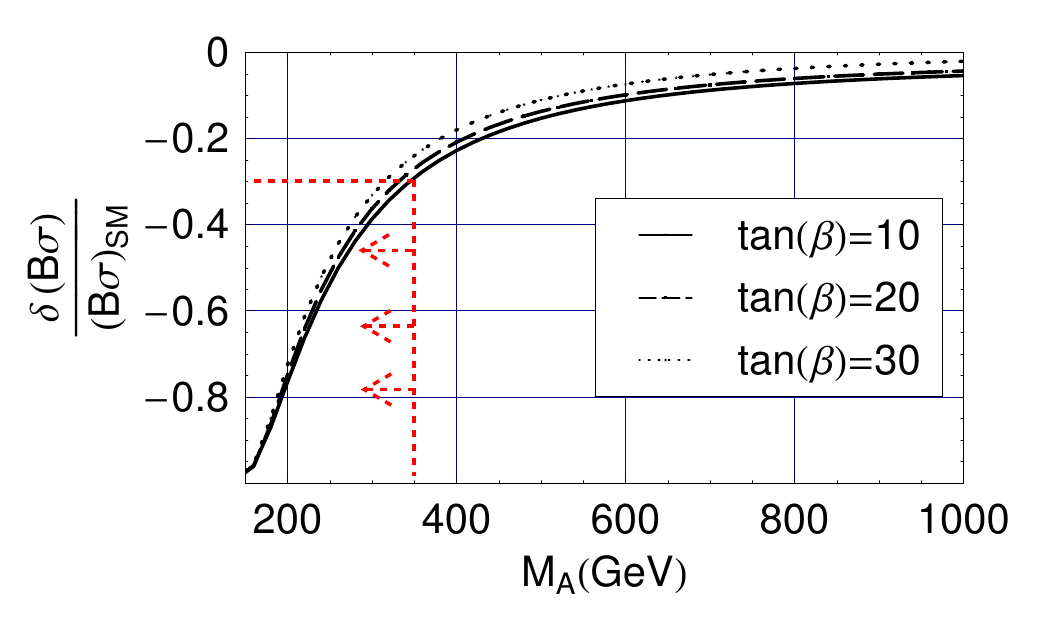}
\caption{The fractional deviation of $\HGG$ as a function
of $M_A$ for $\tan\beta=10, 20$, and 30.
The soft masses are fixed to
evade LHC discovery as $\mu=M_{\tilde{w}}=M_{\tilde{\ell}}=1$ TeV,
$M_{\tilde{g}}=M_{\tilde{Q}}=2.5$ TeV,
$A_t=5\ \mbox{TeV}$, and $A_b=A_{\tau}=0$.
For $M_A<330$ GeV, there is significant deviation
in $\HGG$ from the SM.
}
\label{fig:MSSMTPlot}
\end{center}
\end{figure}

In this work we will restrict our attention
to the range $10<\tan\beta<30$, where both
the top and bottom Yukawa couplings are perturbative
and there is no danger for light or negative s-tau or s-bottom
masses from $\tan\beta$-enhanced mixing
when
$\mu\, m_{b(\tau)}\,\tan\beta\sim M^2_{\tilde{b}(\tilde{\tau})}$,
where $\tilde{b}$ and $\tilde{\tau}$ are respectively
the s-bottom and s-tau.

To study the effects of $M_A$ and $\tan\beta$ in
$\HGG$ in the decoupling limit of heavy sfermions
and gauginos, in Fig.~\ref{fig:MSSMTPlot}
we show the dependence in $\HGG$ as a function of $M_A$
for $\tan\beta=10,20,$ and 30, with other parameters
fixed as
\begin{align}
\mu=M_2=M_{\tilde{\ell}}&=1\ \mbox{TeV},\\
M_{\tilde{g}}=M_{\tilde{Q}}&=2.5\ \mbox{TeV},\\
A_t&=5\ \mbox{TeV},\\
A_b=A_{\tau}&=0.
\label{eq:MSSMTPlot-param}
\end{align}
This set of parameters differs from those in
Eq.~\ref{eq:ScannedSet} because we will eventually
be interested in the MSSM lone Higgs scenarios.
As we will see in the next subsection, these parameters give sufficiently
heavy sfermions and gauginos that are out of the reach at the LHC
with 10 fb$^{-1}$ of integrated luminosity.
We note that the dependence in $\tan\beta$ on $\HGG$ is
small in this range of $\tan\beta$.
In particular, $\HGG$
can be suppressed by more than 30\% for $M_A\lesssim 330$ GeV.
In the next subsection we will investigate whether
this region can be consistent with a lone
Higgs scenario given the expected LHC reach for the
heavy Higgs bosons.

We conclude this subsection with a brief discussion
of the possibility of having an enhancement in
$\HGG$ in the MSSM.
In the lower-right plot of Fig~\ref{fig:MSSMPlot},
in the region with small Higgs mass ($m_h\sim$ 100 GeV)
we have enhanced $\HGG$ that is contrary to
the typical trend of suppression we note earlier.
This is because the enhancement
in the production $\sigma(gg\rightarrow h)$ (due to light s-tops with small mixing)
compensates for the small suppression
in the branching ratio
$\mbox{Br}(h\rightarrow\gamma\gamma)$ (due to large $M_A=1$ TeV).
On the other hand,
in the lone Higgs scenario we consider here with
two heavy s-tops,
the production
cross-section $\sigma(gg\rightarrow h)$
is only enhanced slightly relative to the SM
\cite{Harlander:2004tp},
and the suppression in $\HGG$ is dominantly
due to a suppression in $\mbox{Br}(h\rightarrow\gamma\gamma)$
resulting from $\tan\beta$-enhanced $h\overline{b}{b}$
and $h\overline{\tau}\tau$ couplings, leading
to a larger total decay width of the Higgs boson.
(We also note that it is possible to have a suppression in the
production cross-section $\sigma(gg\rightarrow h)$
in the MSSM, leading
to a so-called gluo-phobic Higgs, when there
is a large hierarchy between the s-tops $m^2_{\tilde{t}_2}\gg
m^2_{\tilde{t}_1}$ and a significant mixing in the s-top sector
\cite{Harlander:2004tp}.)

\subsubsection{Lone Higgs Scenario in the MSSM}
\label{sec:MSSM-LH}
\begin{figure}[h!t]
\begin{center}
\includegraphics [width=5.0in]{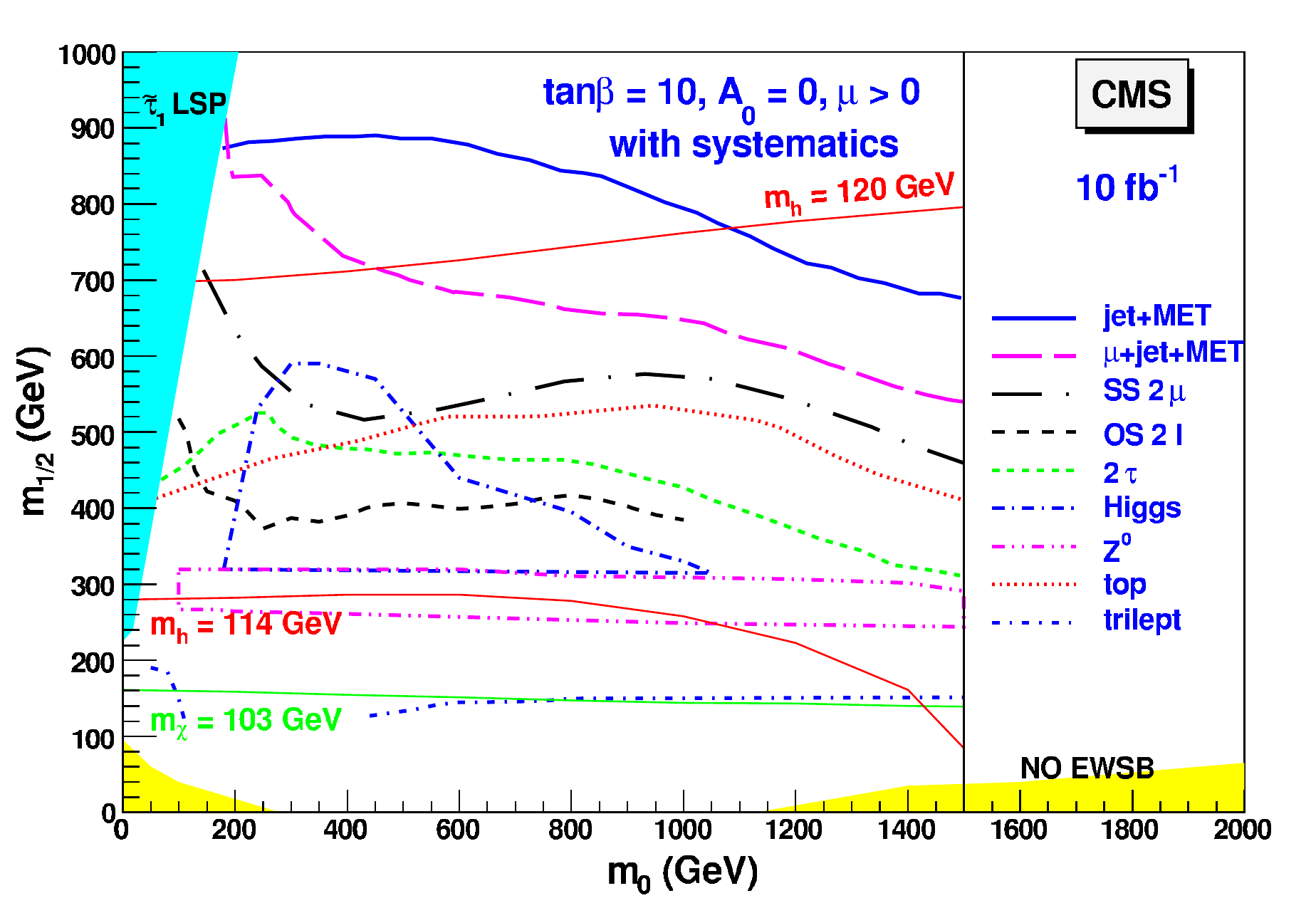}
\caption{
This figure is taken from the CMS TDR \cite{Ball:2007zza},
and summarizes the MSSM reach on $m_0$-$m_{1/2}$ plane
at CMS with an
integrated luminosity of 10 $\invfb$, except
for the Higgs search, which assumes 2 $\invfb$.}
\label{fig:susy_cms}
\end{center}
\end{figure}

To investigate the viable lone Higgs scenarios in
the MSSM, we
here briefly summarize the known results in
the literature regarding the LHC reach of the MSSM
particles.
The summary of MSSM discovery potential at the
CMS experiment at the LHC
for the constrained MSSM (CMSSM) is shown in the CMS TDR~\cite{Ball:2007zza},
and we include it in Fig.~\ref{fig:susy_cms}.
We here focus on reaches of several experimental signatures
that allow us to place general bounds
on the MSSM parameter space, and
we will simplify the MSSM parameter space in
terms of four general types of resonances in addition
to the Higgs boson.
These four types are:
(i) colored superpartners (gluino and squarks),
(ii) sleptons,
(iii) neutralinos and charginos,
and (iv) heavy Higgs bosons.
We discuss the discovery
reach of each in turn.

\paragraph{Gluino and squarks}$\ $\newline
\indent In the CMSSM parameterization
of the MSSM parameters by the set
$\{m_0, m_{1/2}, A_0, \tan\beta, \mbox{sgn}(\mu)\}$,
the colored superpartners are generically
heavier than the non-colored superpartners,
and the search for SUSY involves
tracking down the cascade decays of
pair-produced squarks and/or gluinos.
While the nature of the cascades depends
on the details of the MSSM spectra
(see CMS TDR~\cite{Ball:2007zza} for the various possibilities),
the initial step of the cascade always
involves emitting at least a quark and
all cascades always
end with the stable, lightest superpartner (LSP)
that leaves the detector as missing energy (assuming
that $R$-parity is conserved).
Thus, the signature of jets plus
missing energy provides
the best reach of the MSSM space
(line labelled 'jet+MET' in Fig.~\ref{fig:susy_cms}),
and the reach is roughly $m_{1/2}\sim$ 900 GeV,
corresponding to a gluino mass of
$M_{\tilde{g}}\sim 2.7m_{1/2}\sim$ 2.4 TeV.
We also note that this reach is only mildly dependent on $m_0$,
and the squark masses are less constrained by this reach.
Nevertheless, we make a reasonable simplifying assumption
that the reach for squarks is also $M_{\tilde{Q}}\sim$2.4 TeV.

\begin{figure}[h!t]
\begin{center}
\includegraphics [width=4.0in]{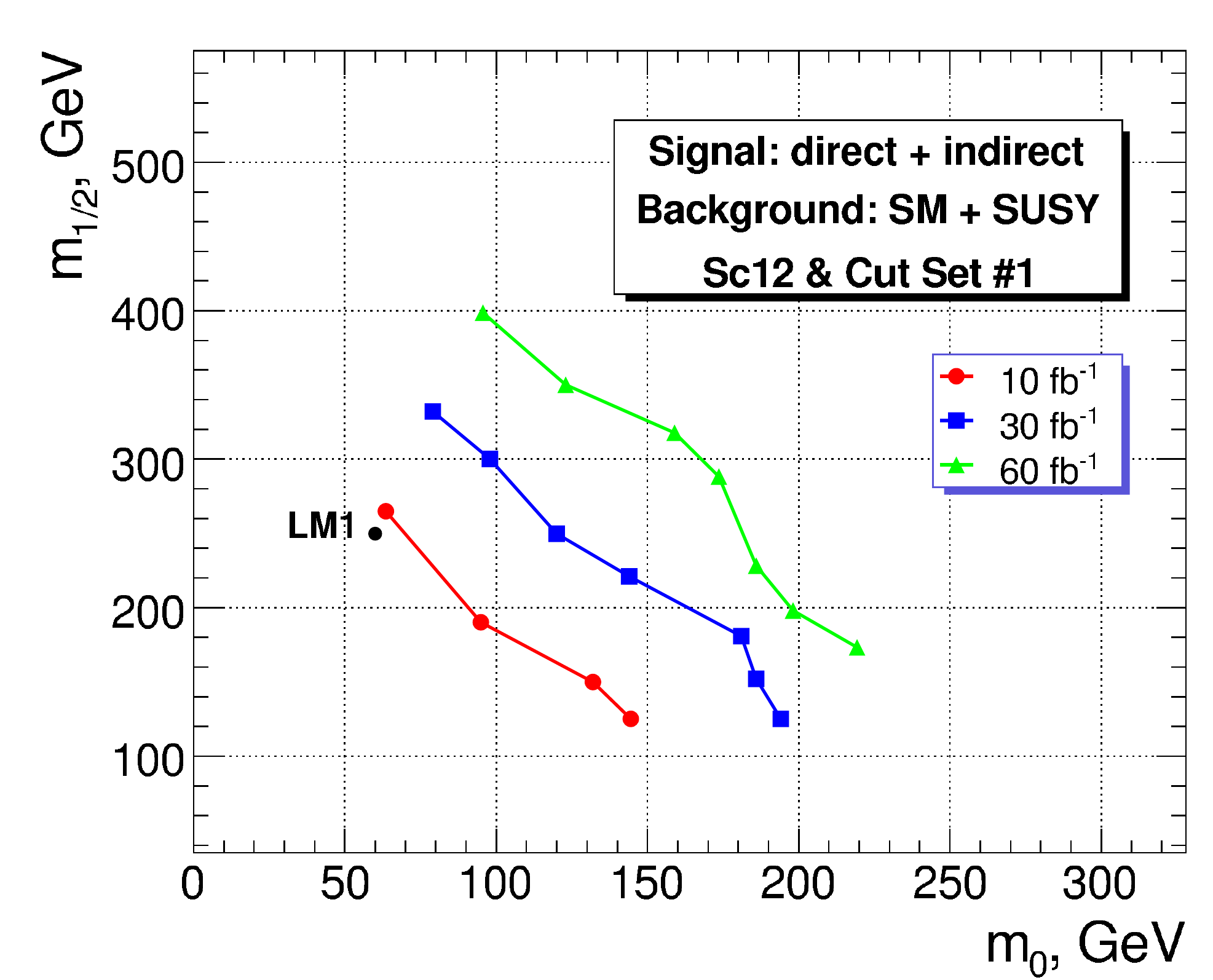}
\caption{
This plot is taken from CMS TDR~\cite{Ball:2007zza} and shows CMS
reach of pair production of slepton
on $m_0$-$m_{1/2}$ plane with $\tan\beta=10$, $A_0=0$,
and the sign $\mu$ is positive.  The experimental signatures of the events
are two leptons, missing energy, and no jets.
}
\label{fig:slepton-bound}
\end{center}
\end{figure}

\paragraph{Sleptons}$\ $\newline
\indent Moving away from the CMSSM parameterization,
if a large hierarchy existed between the colored
superpartners and the non-colored superpartners such that
at the LHC the gluino and squarks can not be directly
produced at the LHC,
then SUSY searches involve the production channels
of non-colored superpartners, such as chargino-neutralino associated
production
and slepton pair-productions.
While the right-handed slepton (superpartner
of the right-handed lepton) can only decay to
the LSP directly in CMSSM, the left-handed slepton
can decay to $\chi_{1,2}^0$ and $\chi_{1}^{\pm}$
if kinematically allowed (depending on $m_{1/2}$).
In both cases of direct slepton pair production
and indirect slepton production (through the decays
of neutralinos and charginos from
associated chargino-neutralino associated
production),
there are always at least two leptons
from the decays of the sleptons.
Furthermore, in the case of direct slepton pair-production,
the two leptons must have opposite sign.
The signature of the direct slepton pair production
and indirect production
then involves two leptons, missing
energy, and jet veto, and
the exclusion plot in the reference is reproduced
in Fig.~\ref{fig:slepton-bound}
(see Chapter 13, Section 15 of the CMS TDR~\cite{Ball:2007zza}
for details on the cuts and significance of
the reach).
For large $\tan\beta$, the largest slepton
mass is roughly
$m^2_{\tilde{l}}\sim m_0^2+(0.5)m^2_{1/2}+0.27 M_Z^2$,
and using $m_0=m_{1/2}=150$ GeV from
Fig.~\ref{fig:slepton-bound}, we obtain
$m_{\tilde{l}}\sim 200$ GeV.
For further references, there are detailed studies of
slepton searches in the literature
\cite{delAguila:1990yw}\cite{Baer:1993ew}\cite{Baer:1995nq}.
%

\paragraph{Neutralinos and charginos}$\ $\newline
\indent The reach of neutralino($\chi_i^0$)/chargino($\chi^{\pm}_i$)
sector involves a tri-lepton signature that arises
from $pp\rightarrow \chi_2^0\chi_1^{\pm}X$ with
$\chi_2^0\rightarrow \ell^{+}\ell_{-}\chi_1^0$
(through a slepton $\tilde{\ell}$, which may be real or virtual)
and
$\chi_1^{\pm}\rightarrow\chi_1^{0}W^{(*)\pm}\rightarrow\chi_1^{0}\ell^{\pm}\nu_{\ell}$.
The lepton pair from the decay of $\chi_2^{0}$ will be of the same flavor, opposite sign
(SFOS),
while the lepton from $\chi_1^{\pm}$ can be any flavor.
In addition to the three leptons and missing energy,
no jets participate in this process and a jet veto
can be used
(see Chapter 13, Section 14 of the CMS TDR~\cite{Ball:2007zza} for details),
and the reach is presented in the bottom blue curve in Fig.~\ref{fig:susy_cms}
denoted as ``trilept''.
As with the case of the gluino/squark reach,
we see that the curve is only mildly dependent
on $m_{0}$ and has a value of $m_{1/2}\sim 150$ GeV
for $m_0>400$ GeV, and $m_{1/2}$ fluctuates between
100 GeV and 200 GeV for $m_0<400$ GeV.
The wino (bino) mass is related to $m_0$
in CMSSM by $M_2\sim 0.8 m_{1/2}$
($M_1\sim 0.4 m_{1/2}$).
Thus, taking the conservative reach of $m_{1/2}\sim$ 200 GeV
gives $M_2\sim 160$ GeV and $M_1\sim 80$ GeV.
To obtain the full spectra in
the neutralino/chargino sector requires the
knowledge of the Higgsino mass $\mu$.
However, assuming that $\mu>M_2$ and that
$\mu$ is not nearly degenerate with
$M_2$, we do not expect large mixing
in the neutralino/chargino sector,
and we can approximate $M_{1,2}$ as the masses
of $M_{\chi_{1,2}^0}$

\begin{figure}[h!t]
\begin{center}
\includegraphics [width=3.125in]{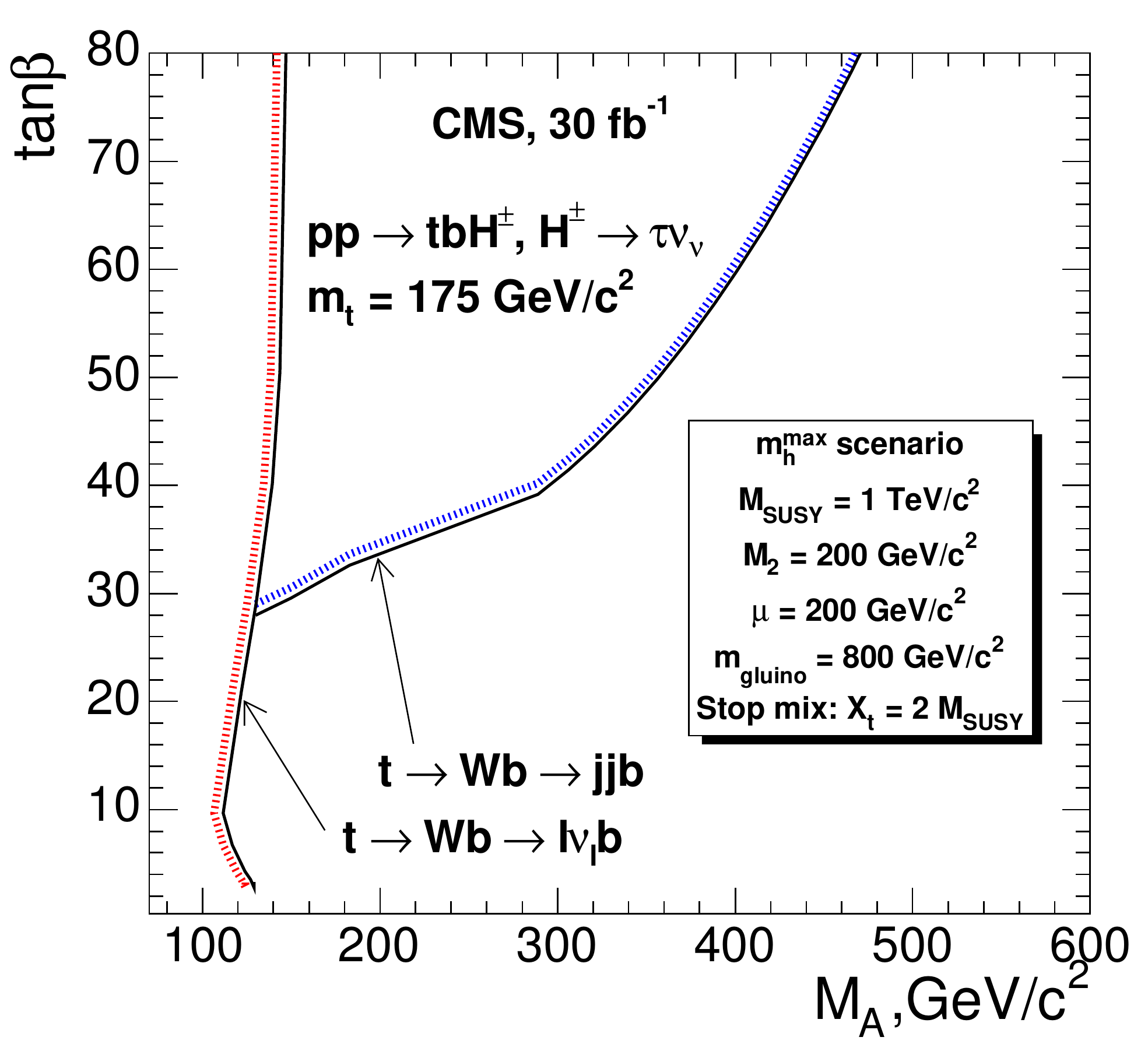}
\includegraphics [width=3.000in]{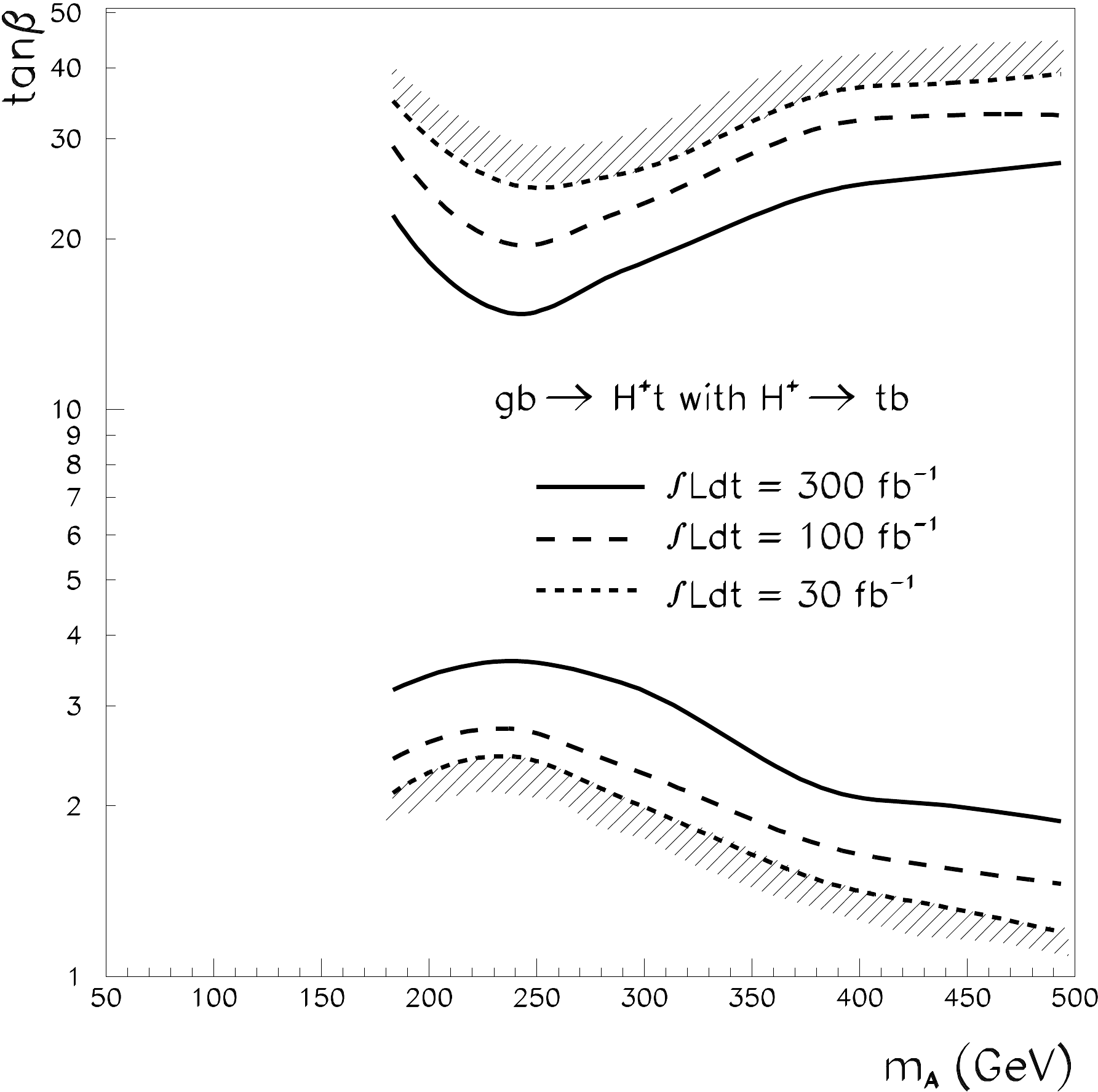}
\caption{
The left plot is taken from the CMS TDR \cite{Ball:2007zza},
and the right plot taken from the ATLAS TDR \cite{:1999fr}.
These plots summarizes the LHC reach of heavy
charged Higgs bosons at the CMS (left) and
ATLAS (right) using channels as indicated on the plots.
}
\label{fig:ChargeHiggs}
\end{center}
\end{figure}

\begin{figure}[h!t]
\begin{center}
\includegraphics [width=3.125in]{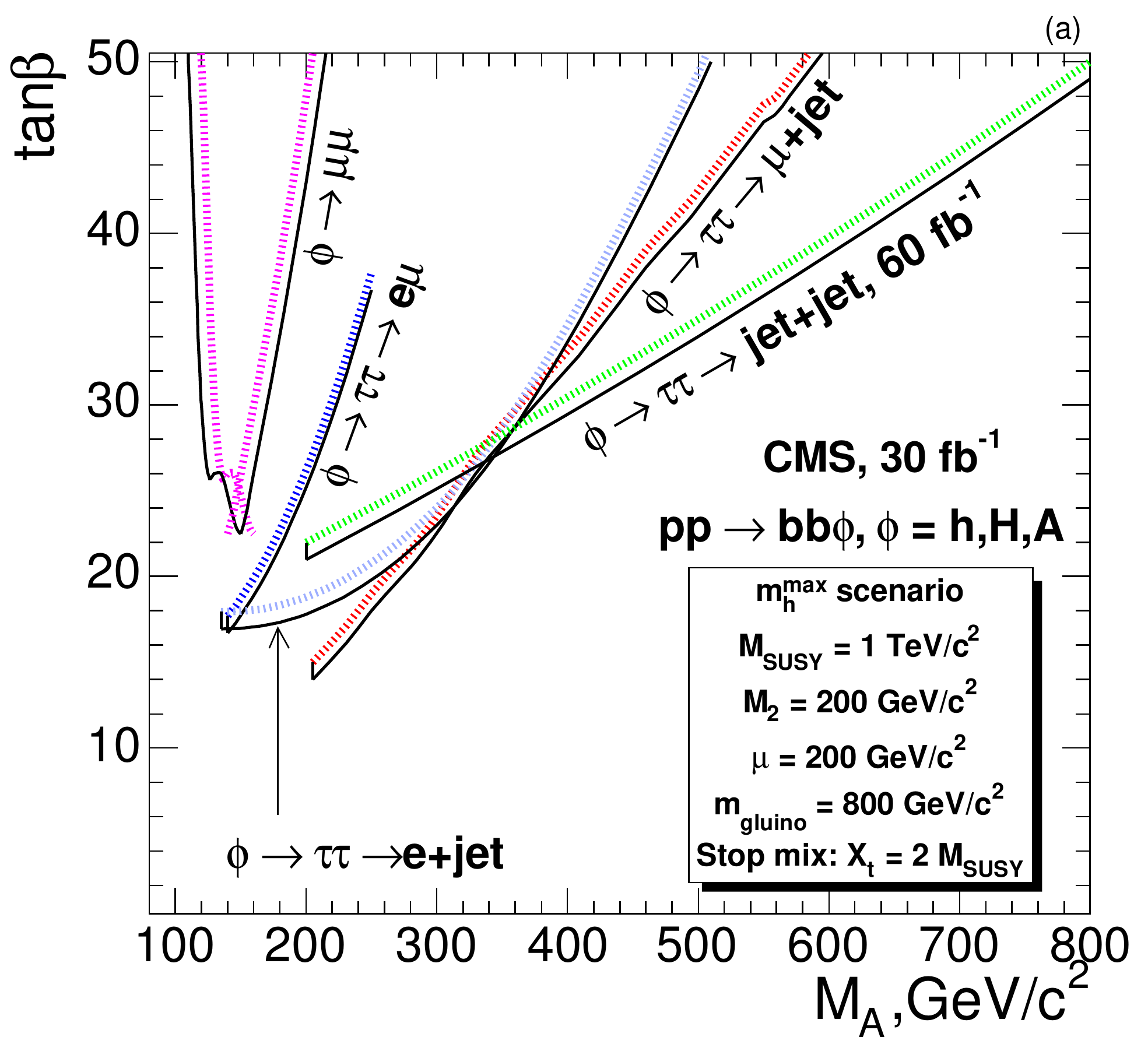}
\includegraphics [width=3.125in]{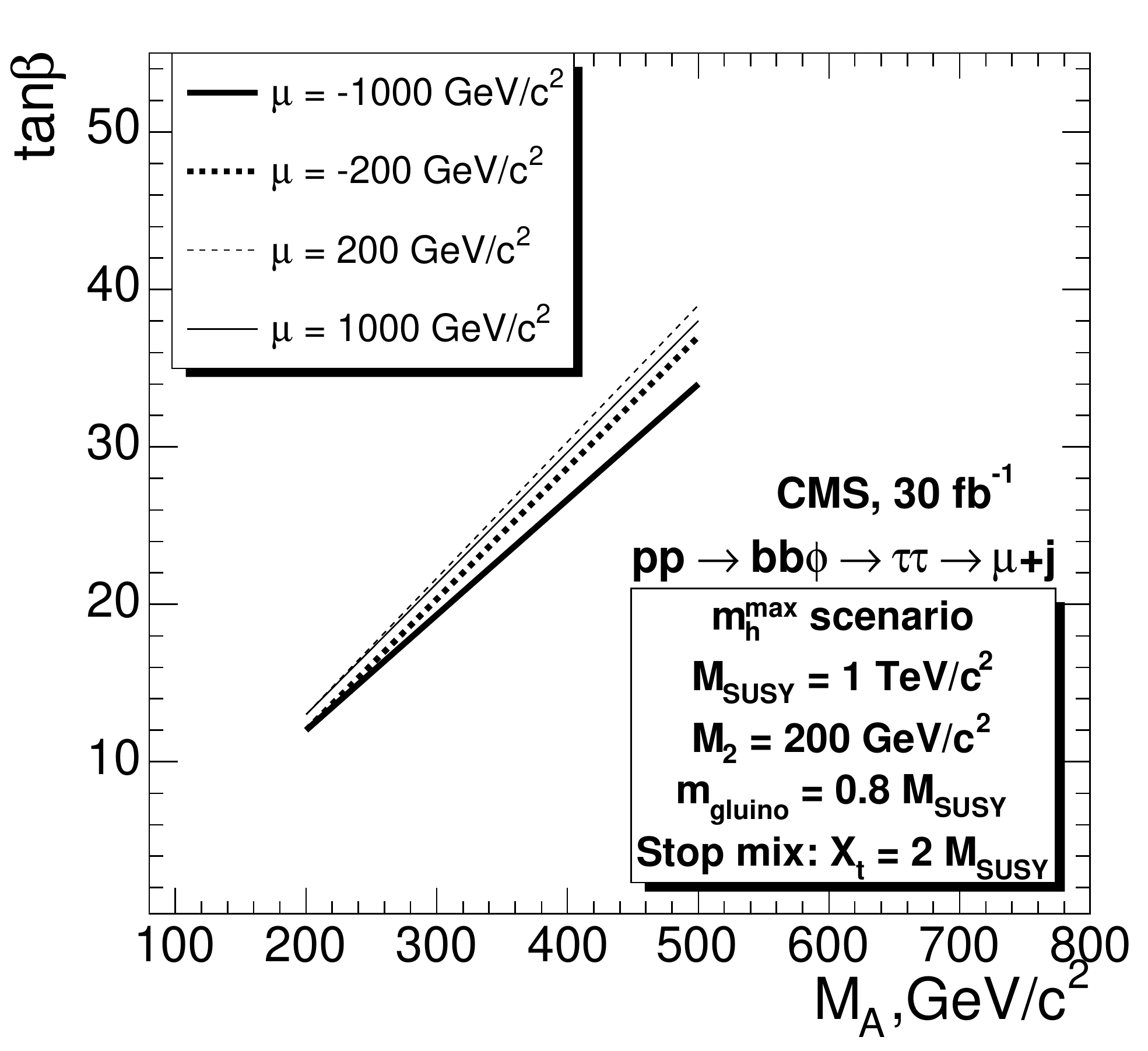}
\includegraphics [width=3.0in]{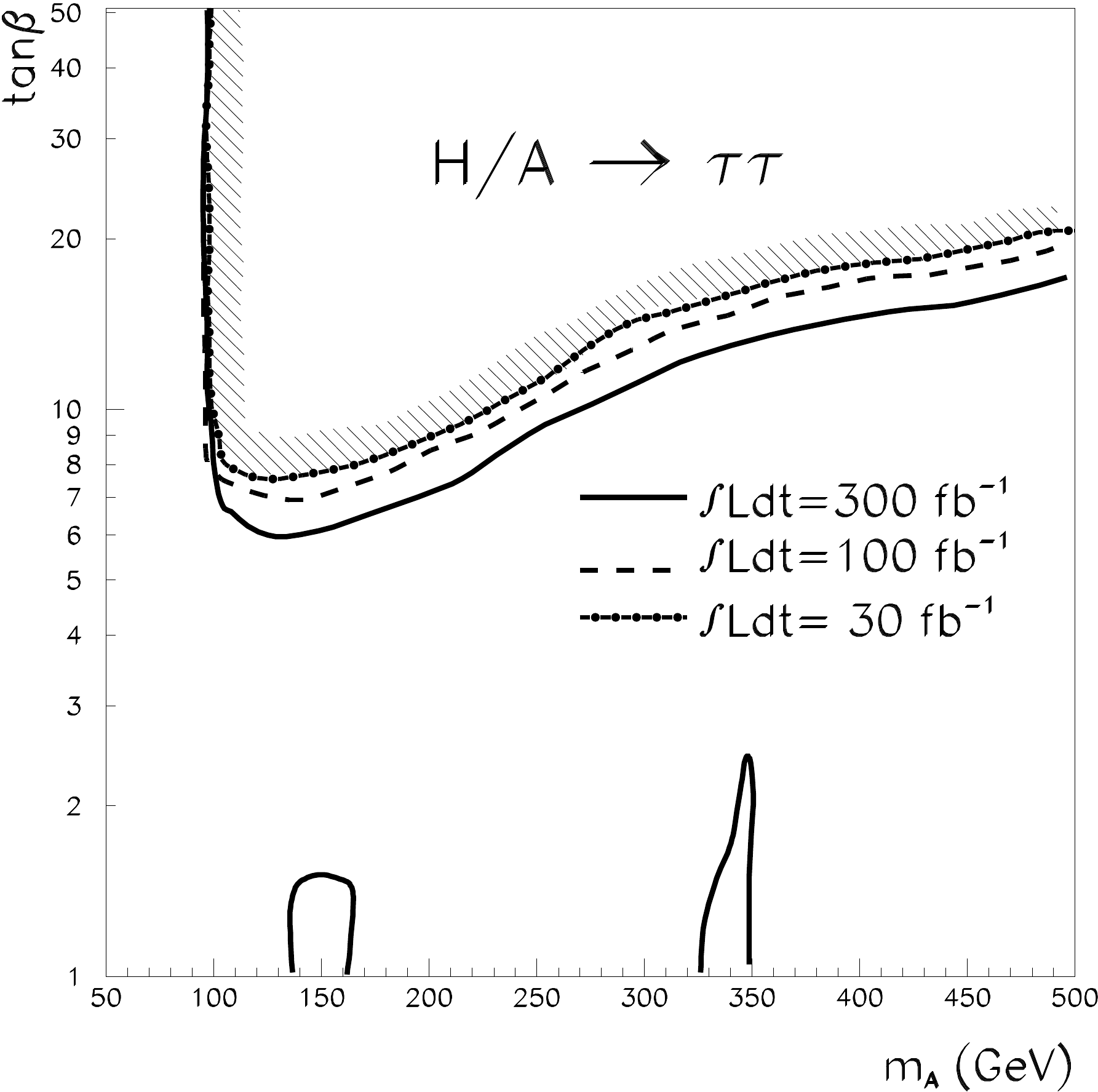}
\includegraphics [width=3.0in]{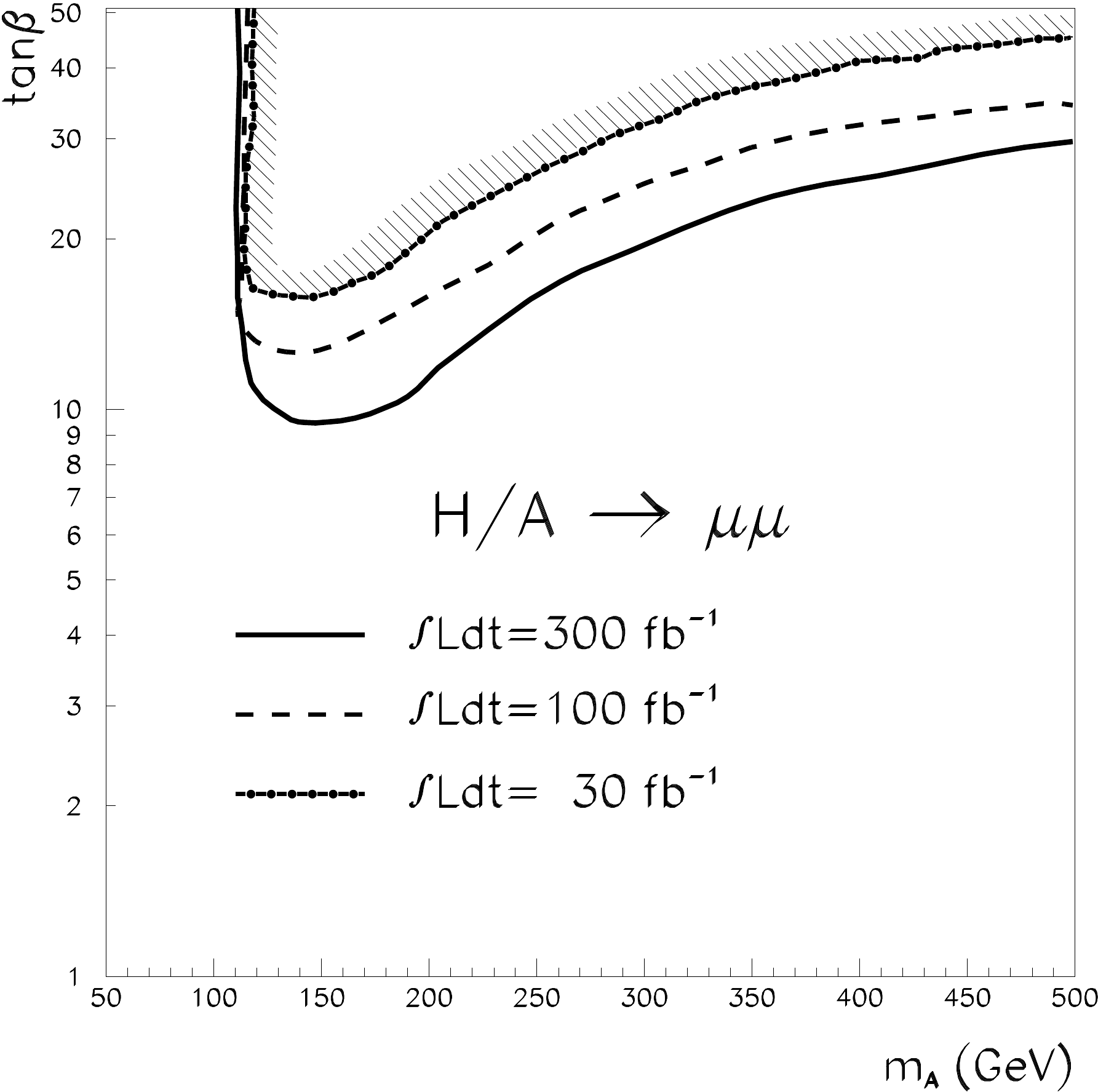}
\caption{
Fig.~\ref{fig:NeutralHiggs}(a) is taken from the CMS TDR \cite{Ball:2007zza},
Fig.~\ref{fig:NeutralHiggs}(b) from Gennai et al.~\cite{Gennai:2007ys},
and Figs.~\ref{fig:NeutralHiggs}(c) and (d) from ATLAS TDR \cite{:1999fr}.
These plots summarizes the CMS and ATLAS reach of heavy
neutral Higgs bosons in various channels,
as indicated on the plots.
Fig.~\ref{fig:NeutralHiggs}(c) sets the strongest
constraints on having a lone Higgs scenario with
large deviations in $\HGG$.
}
\label{fig:NeutralHiggs}
\end{center}
\end{figure}

\paragraph{Higgs bosons}$\ $\newline
\indent Because the variation in $\HGG$ in
the MSSM has strong dependence on $M_A$ in the decoupling limit
of heavy s-fermions and gauginos, the LHC reaches of
the heavy Higgs bosons play important roles in
the determination of whether we have
a lone Higgs scenario with a large deviation
in $\HGG$.
The reach of the MSSM heavy Higgs bosons has been studied
in the literature
\cite{:1999fr}\cite{Ball:2007zza}\cite{Gennai:2007ys},
and we have taken figures from these references
in Figs.~\ref{fig:ChargeHiggs} and \ref{fig:NeutralHiggs}.

Although in this work we focus on the lone Higgs scenario
with 10 fb$^{-1}$ of LHC data, we could only find discovery
reach for the heavy Higgs bosons with 30 fb$^{-1}$ of data.
We will assume that the reach of the MSSM heavy Higgs bosons
are further with 30 fb$^{-1}$ of data than with 10 fb$^{-1}$,
and conservatively apply the discovery reach of the heavy MSSM Higgs bosons
at 30 fb$^{-1}$ to our 10 fb$^{-1}$ lone Higgs scenario.

The discovery reach of the charged Higgs bosons
$H^{\pm}$ is presented Fig.~\ref{fig:ChargeHiggs},
with $pp\rightarrow tbH$, where $H^{\pm}$ is
either produced through
the decay of a top quark
$t\rightarrow H^{+}b$ produced in $\overline{t}t$ production
if $M_{H^{\pm}}<m_t$, or, if $M_{H^{\pm}}>m_t$, produced
through a virtual $b^{\ast}$ via
$b^{\ast}\rightarrow t H^{-}$ in a $\overline{b}b$
production.
The charged Higgs boson
then decays via $H^{-}\rightarrow\tau
\overline{\nu}_{\tau}$, and as the coupling of $H^{+}\overline{t}b$ is
$\tan\beta$-enhanced,
stronger bounds can be placed
with larger $\tan\beta$.
However, for $\tan\beta\lesssim 30$,
the reach on $M_{A}$
is less than $M_{A}=200$ GeV with CMS,
while ATLAS does not set a bound
on $M_A$ for $4\leq\tan\beta\leq 25$.
Thus, the discovery reach for
the charged Higgs bosons does not impose
a severe constraint on $M_A$.

The discovery reaches of
the neutral Higgs bosons $h, H$, and $A$
at CMS and ATLAS are
presented Fig.~\ref{fig:NeutralHiggs}
using various processes
as indicated on the plots.
As with the charged Higgs, these
searches depend on $\tan\beta$-enhanced
couplings.
For a given value of $M_A$,
these heavy Higgs boson
discovery reaches limit the lone Higgs scenario
with an upper bound on $\tan\beta$.
The strongest discovery reach for
the neutral Higgs boson comes from the ATLAS search
involving $H/A\rightarrow\overline{\tau}\tau$
as shown in Fig.~\ref{fig:NeutralHiggs}(c),
which places an upper bound
of $\tan\beta < 15$ if we are to have $M_A<330\ \mbox{GeV}$
in a lone Higgs scenario.
For larger values of $\tan\beta$, we would need
larger values of $M_A$ to have viable lone Higgs scenarios,
and it is difficult to distinguish the lone Higgs scenario
from the SM by the suppression in $\HGG$.
%
With 300 fb$^{-1}$ of integrated luminosity,
the LHC can also discover the $A$ if $\tan\beta$ is small
($1\leq\tan\beta\leq 2$), as shown in Fig.~\ref{fig:NeutralHiggs}(c), through
the process $gg\rightarrow A\rightarrow \overline{\tau}\tau$,
whose rate is significantly larger than
that of the SM Higgs boson $gg\rightarrow h\rightarrow \overline{\tau}\tau$.
%
As the bounds from ATLAS are more stringent,
we will mainly use Fig.~\ref{fig:NeutralHiggs}(c) to
set bounds on the lone Higgs scenario in the MSSM
in the next section.
(For details, see the ATLAS TDR\cite{:1999fr}, the CMS TDR \cite{Ball:2007zza},
Gennai et al.~\cite{Gennai:2007ys},
and references therein.)
%

\paragraph{Possibilities of a Lone Higgs Scenario in MSSM} $\ $\newline

\begin{table}[h!t]
\begin{center}
\caption{Summary of LHC reach for
different types of superpartners of
the MSSM with 10 $\invfb$ of integrated
luminosity.  The reach of the heavy Higgs bosons
involve 30 $\invfb$ of integrated luminosity.
The assumptions
we make are in parenthesis.}
\label{tb:susy-summary}
\vspace{0.125in}
\begin{tabular}{|l|l|l|}
\hline
Superpartner  &  LHC reach   & Signature used \\ \hline\hline
Gluino/squarks & $M_{\tilde{g}}\sim 2.5$ TeV
($M_{\tilde{q}}\sim 2.5$ TeV) & Jets and missing energy \\ \hline
Sleptons & $M_{\tilde{\ell}}\sim 200$ GeV & OS leptons, missing
energy, and jet veto\\ \hline
Neutralinos/charginos & $M_{2}\sim 160$ GeV ($\mu>M_2$)&
Tri-lepton, missing energy, and jet veto\\ \hline
Heavy Higgs bosons & $M_{A}\sim 230(480)$ GeV ($\tan\beta=10(20)$)&
$\phi\rightarrow\tau^{+}\tau^{-},\tau\rightarrow(jj,ej,\mu j)$  \\ \hline
\end{tabular}
\end{center}
\end{table}

\indent The LHC reach for the four types of
superpartners is summarized in Table \ref{tb:susy-summary}.
To obtain a consistent lone Higgs scenario with the MSSM,
we simply require the superpartners to be heavier
than the discovery reach at the LHC.
On the other hand, we see from Figs.~\ref{fig:MSSMPlot}
and \ref{fig:MSSMTPlot} that
the signal $\HGG$ is most sensitive to $M_A$.
With all s-fermions and gauginos out of reach at 10 $\invfb$,
having a large deviation in $\HGG$ requires $M_A<330$ GeV
(see Fig.~\ref{fig:MSSMTPlot}),
which
is only consistent with a lone Higgs scenario
for $\tan\beta\lesssim 15$ from Fig.~\ref{fig:NeutralHiggs}(c).

This is a very strong constraint: with our self-imposed
range of $10<\tan\beta<30$,
with 10 $\invfb$ of LHC data,
if we have only discovered a Higgs boson whose deviation
in $\HGG$ is suppressed by more than 30\% relative to
the SM, then it can be consistent with the MSSM only
if $M_A<330$ GeV and $\tan\beta< 15$.
Furthermore, given $\tan\beta$ ($M_A$)
we can set more stringent bounds on $M_A$ ($\tan\beta$)
using the upper-right plot in Fig.~\ref{fig:NeutralHiggs}(c).
Suppose we know that $\tan\beta=10$, for example, from the
branching ratio of $h\rightarrow\overline{\tau}\tau$.
In this case, the upper bound on $M_A$ is $M_A < 230$ GeV because
any lower value of $M_A$ would give rise to additional resonances
at the LHC.

\begin{figure}[h!t]
\begin{center}
\includegraphics [width=4.25in]{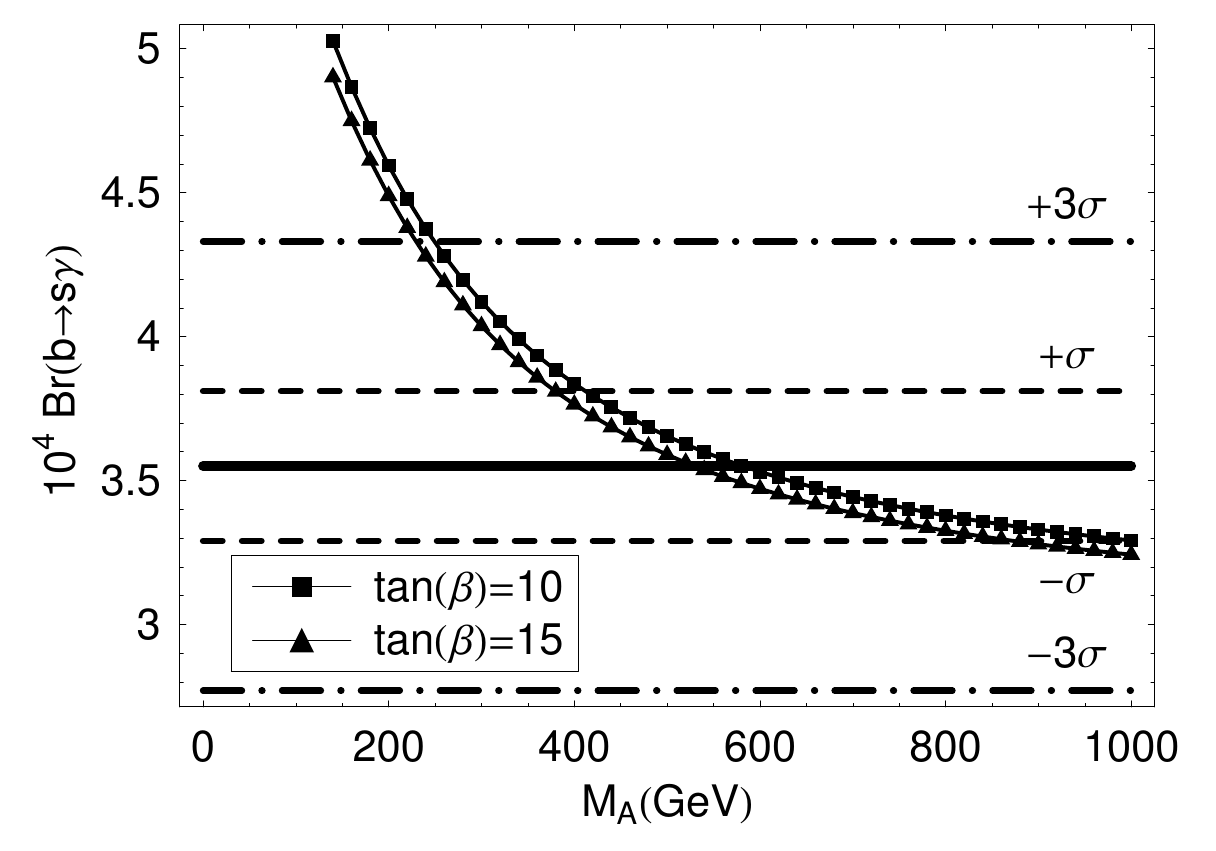}
\caption{
The branching ratio $\mbox{Br}(b\rightarrow s\gamma)$
as a function of $M_A$ for $\tan\beta=10$ and $\tan\beta=15$.
The other parameters are fixed as in Eq.~(\ref{eq:MSSMTPlot-param})
to evade the discovery of the s-fermions, the charginos,
and the neutralinos.
The horizontal lines indicate current experimental value
(solid), 1$\sigma$ uncertainties (dashed), and
3$\sigma$ uncertainties (dot-dashed).
}
\label{fig:BSGPlot}
\end{center}
\end{figure}

We can further constrain viable MSSM lone Higgs scenarios with
a large deviation in $\HGG$ from
the measured value of the branching
ratio $\mbox{Br}(b\rightarrow s\gamma)$ with 1$\sigma$ uncertainty
\cite{Barberio:2006bi}
\begin{align}
\mbox{Br}(b\rightarrow s\gamma)_{\smbox{exp}}=(355\pm 26)\times 10^{-6},
\label{eq:BSGexp}
\end{align}
if we assume
minimal flavor violation (MFV) scenario.
A rigorous definition of MFV can be
found in D'Ambrosio~\cite{DAmbrosio:2002ex}.
In the MSSM we can have MFV with universal soft s-fermion masses and
having the trilinear couplings proportional to the Yukawa couplings at
an arbitrary energy scale, which we pick to be near the weak scale.
In Fig.~\ref{fig:BSGPlot}, we plot $\mbox{Br}(b\rightarrow s\gamma)$
using \verb"SusyBsg" \cite{Degrassi:2007kj} (with the aid
of \verb"SOFYSUSY"~\cite{Allanach:2001kg}) as a function
of $M_A$ for $\tan\beta=10$ and $\tan\beta=15$, holding
all other parameters fixed as in Eq.~(\ref{eq:MSSMTPlot-param}).
With these parameters, the s-fermions, charginos, and neutralinos
are all heavy enough to evade
discovery at the LHC with 10 fb$^{-1}$.
From Fig.~\ref{fig:BSGPlot}, for $\tan\beta=10$ (15), we
must have $M_{A}>420$ GeV (380 GeV) to be consistent with
with $\mbox{Br}(b\rightarrow s\gamma)$ within 1$\sigma$ uncertainties,
and this excludes a lone Higgs scenario with a large suppression
compared to the SM prediction since such lone Higgs scenario requires $M_A<350$ GeV
(see Fig.~\ref{fig:MSSMTPlot}).
On the other hand, to be consistent with
$\mbox{Br}(b\rightarrow s\gamma)$ within 3$\sigma$ uncertainties,
for $\tan\beta=10$ (15), the limits
on $M_A$ relaxes to  $M_{A}>260$ GeV (240 GeV),
and we can have lone Higgs scenarios with large suppressions
in $\HGG$ compared to the SM prediction.
If the LHC discovers a lone Higgs scenario
with a suppression in $\HGG$ of more than 55\% from the SM,
then Fig.~\ref{fig:MSSMTPlot} implies that $M_A< 240$ GeV,
which is not consistent
with the measurement of $\mbox{Br}(b\rightarrow s\gamma)$ assuming MFV.
In this case, if the MSSM is to explain the observed suppression,
the flavor structures in the MSSM must deviate from
those given by MFV.
Since we are interested mainly in
constraints on lone Higgs scenarios from direct
searches and the assumption of MFV imposes very stringent constraints,
we do not assume MFV in our current work.
%

As we will show a later subsection,
the LHT model can
also give rise to a lone Higgs scenario with
a large suppression in $\HGG$ relative to
the SM, and we will investigate how we may distinguish
between these two models in Section~\ref{sec:HGG-lone}.


\subsection{Littlest Higgs model with $T$-parity}
\subsubsection{$\HGG$ in LHT}
\label{sec:LHT-HGG}
In Littlest Higgs model with $T$-parity
\cite{ArkaniHamed:2002qy}\cite{Cheng:2003ju}\cite{Cheng:2004yc}\cite{Low:2004xc},
based on little Higgs models
\cite{ArkaniHamed:2002qx}\cite{noteLHT}\cite{Han:2003wu}\cite{Schmaltz:2005ky},
the SM Higgs doublet is
a pseudo Nambu-Goldstone boson (NGB) of
two independent spontaneously broken symmetries
at a scale $\Lambda$.
The collective symmetry breaking mechanism of generating the Higgs mass
ensures that its quadratic divergence
vanish at one-loop level, and the electroweak
scale can be stabilized with $\Lambda\sim$ 10 TeV.
The contributions to electroweak precision observables are
loop-suppressed
with the introduction of a $T$-parity.
Most of the new states that are accessible at the LHC
must be pair-produced because they are odd under $T$-parity while
all the SM particles are even under $T$-parity.
The mass scale of these $T$-odd particles is of the order
$f\sim (4\pi)^{-1}\Lambda$, and the lower bound
on $f$ from electroweak precision tests is about 500 GeV \cite{Hubisz:2005tx}.

\begin{figure}[h!t]
\begin{center}
\includegraphics [width=3.15in]{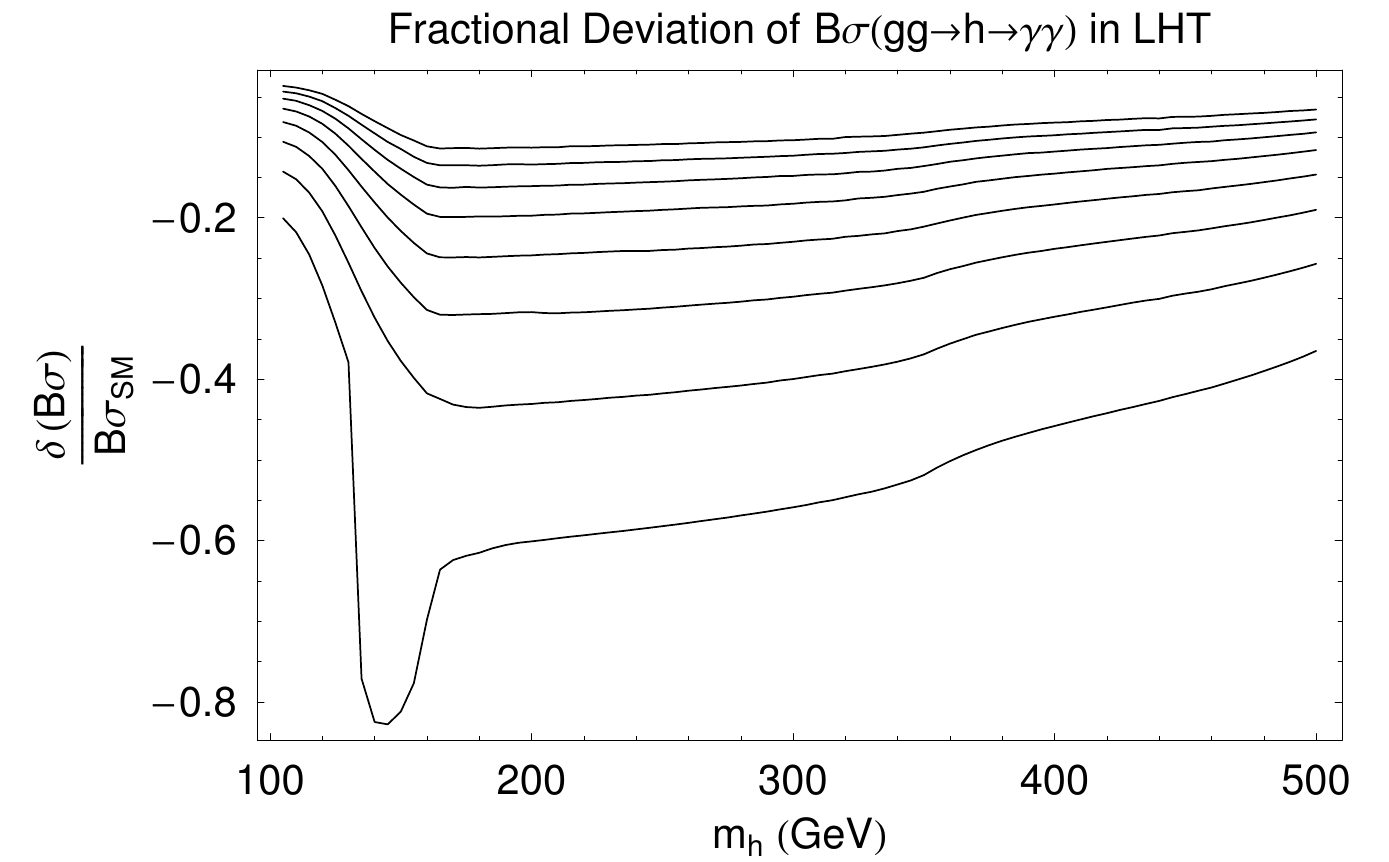}
\includegraphics [width=3.15in]{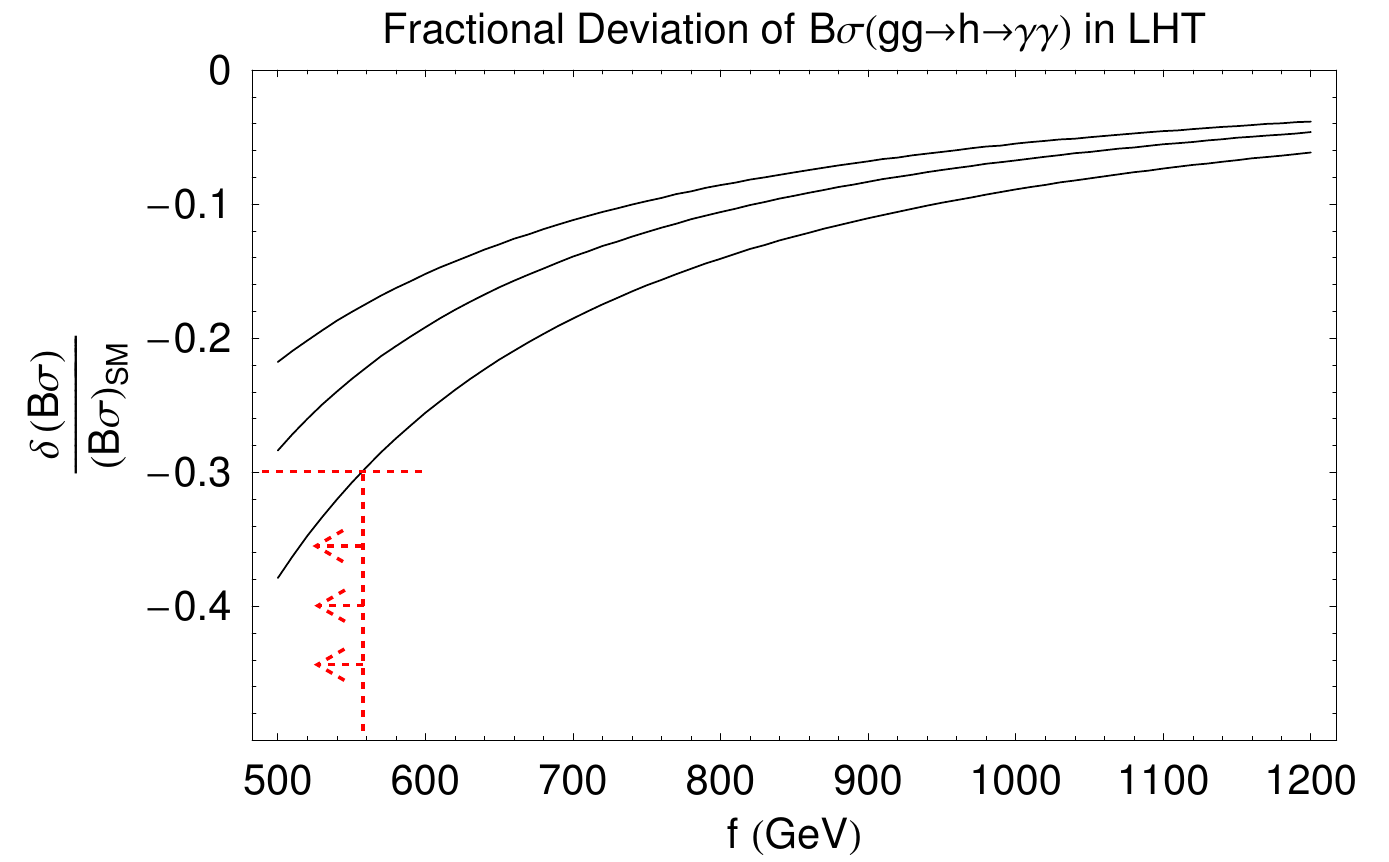}
\caption{The plot on the left shows the fractional deviation of $\HGG$ as a function
of Higgs mass in LHT for various values of $f$=
(500, 600, 700, 800, 900, 1000, 1100, 1200) GeV.
The lowest curve on the plot (showing the most deviation
from the SM) correspond to $f$=500 GeV.
The plot on the right shows the fractional deviation of $\HGG$ as a function
of $f$ mass in LHT for $m_h$=110 (top curve), 120 (middle), and 130 (bottom) GeV.
For $f<560$ GeV, the suppression is greater than 30\%,
and significantly different from the SM.}
\label{fig:LHTPlot}
\end{center}
\end{figure}

The Higgs production and di-photon branching ratio
in LHT has been studied in Chen et al.~\cite{Chen:2006cs}.
As noted in the reference,
the gluon-fusion production cross section in LHT is always suppressed
relative to the Standard Model.
The reference also points out that
the gluon fusion process $gg\rightarrow h$
is only dependent on $f$,
and independent of the parameters of the extended top sector
because changes in
the masses of the top-partners are compensated
by changes in the $h\overline{T}_{+}T_{+}$
coupling, where $T_{+}$ is the $T$-even top partner.
The di-photon branching ratio of the Higgs
boson, however, can be enhanced even though the di-photon width is smaller than
that of SM.
In the left (right) plot in Fig.~\ref{fig:LHTPlot},
we show the deviation in $\HGG$ from the SM
values for several values of $f$ ($m_h$) as a function
of $m_h$ ($f$).
The decay channel $h\rightarrow A_{-}A_{-}$ is allowed
when $m_h>2 M_{A_{-}}$, and for low values of
$f=500$ GeV, this gives a sharp drop in the suppression of $\HGG$ relative to the SM
around $m_h\sim 150$ GeV.
We first note that, similar to the case of heavy s-fermions and gauginos in the MSSM,
the signal is always suppressed relative to the SM,
and this may give us two interpretations
of a lone Higgs scenario with a large suppression in $\HGG$.
We also note that, from Fig.~\ref{fig:LHTPlot}, for $m_h<120$ GeV,
the signal deviates less than 30\% from the SM,
and may not be distinguished
from the SM.

To have large deviation in $\HGG$ in LHT,
we then need both $m_h>120$ GeV as well
as a low value of $f$.
For example, with $m_h=130$ GeV,
the signal only deviates more than
30\% from the SM for $f<560$ GeV.
To investigate whether
this
is consistent with the lone Higgs scenario,
we have to first summarize the discovery
potential of the various new resonances in LHT,
which we turn to next.
%

\subsubsection{Lone Higgs Scenario in LHT}
\label{sec:LHT-LH}
The discovery reaches for the $T$-odd particles
have been studied in the literature, and
we discuss in turn the three classes of new resonances of LHT that
can potentially be seen at the LHC:
(i) $T$-odd gauge bosons ($W^{\pm}_{H}$ and $A_{H}$),
(ii) $T$-odd quarks ($Q_{-}$) and leptons ($L_{-}$)
that are $T$-partners to
the SM quark and leptons of the first two generations,
and,
(iii) $T$-odd and $T$-even top quarks $(T_{\pm})$.
%

\begin{figure}[hpt]
\begin{center}
\includegraphics [width=5.5in]{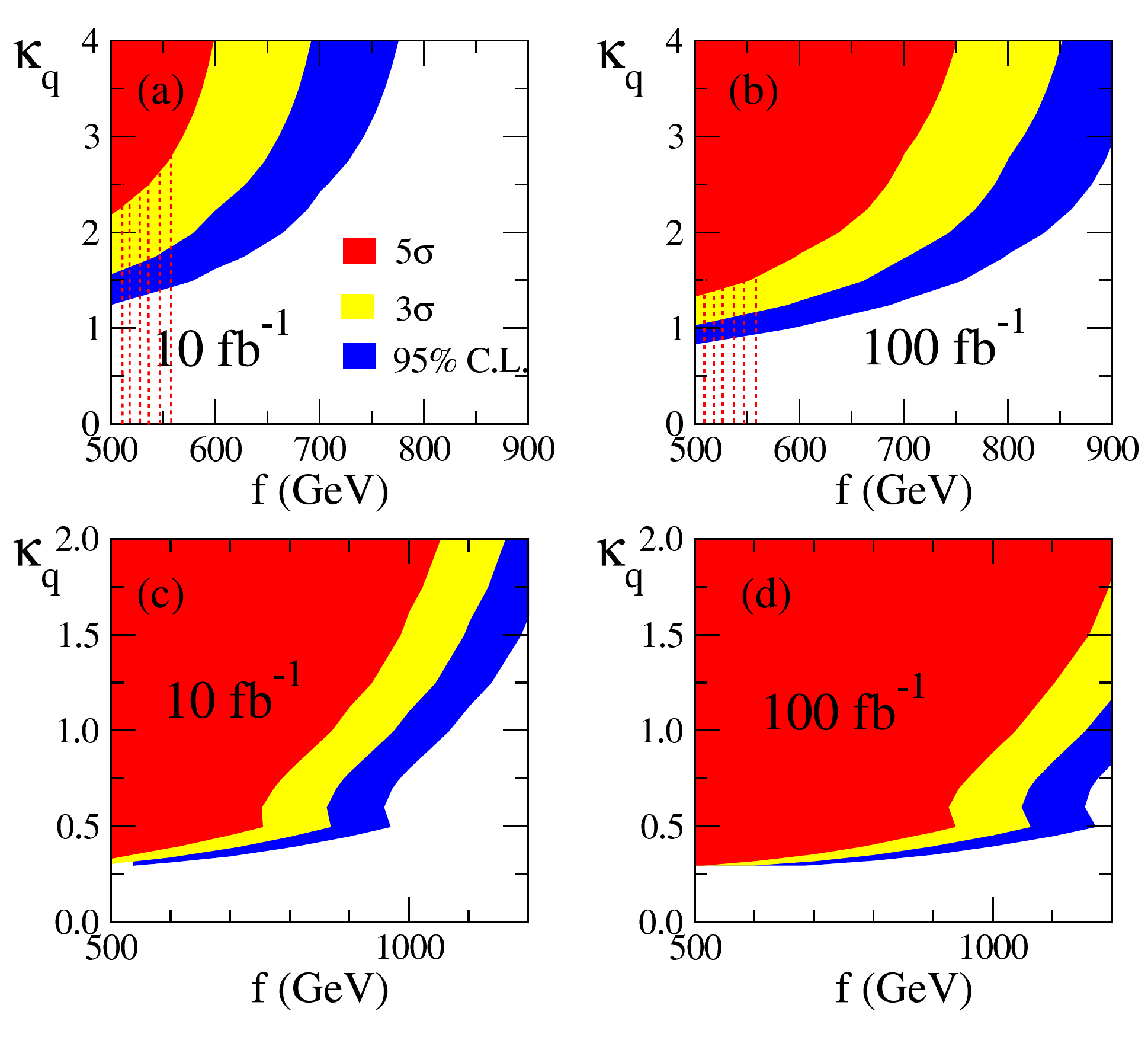}
\caption{These plots are taken from Cao et al.~\cite{Cao:2007pv} and
present the discovery contours of $T$-odd gauge bosons in on
$\kappa_q$-$f$ plane.
The top two plots have $\kappa_{\ell}=0.5$,
so that $W_{H}\rightarrow W A_{H}$ is the
only allowed decay channel at tree-level.
The
lower two plots have $\kappa_{\ell}=0.3$,
so $W_{H}$ dominantly decays through
the cascade
$W_{H}\rightarrow L_{-}\ell\rightarrow A_{H}\ell\ell$.
For the case of $\kappa_{\ell}=0.5$, we have
hashed with vertical lines
the regions viable with a lone Higgs scenario
for a large deviation in $\HGG$.}
\label{fig:LHT-potential}
\end{center}
\end{figure}

\paragraph{$T$-odd gauge bosons}$\ $\newline
\indent The LHC phenomenology of the $T$-odd gauge bosons
has been studied by Cao et al.~\cite{Cao:2007pv},
and the discovery reach of its results are presented in
Fig.~\ref{fig:LHT-potential}.
While the masses of the $T$-odd gauge bosons
are determined by $f$ alone ($M_{W_H}=M_{Z_H}=g f\sim
0.64 f$ and $M_{A_H}=g^{\prime}(\sqrt{5})^{-1}f\sim 0.16f$),
the masses of the $T$-odd fermions for the first and second generations
involve additional inputs $\kappa_q$ and
$\kappa_\ell$, and are related to $f$ by
$M_{(Q_{-},L_{-})}\sim\sqrt{2}\kappa_{(q,\ell)} f$.
The production of $T$-odd
gauge bosons $\overline{q}q\rightarrow W^{+}_{H}W^{-}_{H}$
is dominated by $s$-channel
exchange in $Z^{\ast}$ and the $t$-channel exchange in $Q^{\ast}_{-}$
interferes destructively with the $Z^{\ast}$-exchange
diagram.
By raising $\kappa_q$, the $Q_{-}$-exchange amplitude
becomes smaller, and the production
cross section is enhanced, leading to a stronger
reach for $W^{\pm}_{H}$.

The search strategies for discovering $W_{H}$ depends
crucially on whether the channels
$W^{\pm}_{H}\rightarrow \ell L_{-}, q Q_{-}$
are kinematically accessible, and as such depends
on $\kappa_q$ and $\kappa_{\ell}$.
Assuming $\kappa_q>1$ and $\kappa_{\ell}=0.5$ so
that $W_{H}\rightarrow WA_{H}$ is the only
decay channel of $W_{H}$, the signal of pair-production
of $W_{H}W_{H}$
then includes the leptonic decays of $W$
and missing energy because $A_{H}$ is stable.
From the top two plots of Fig.~\ref{fig:LHT-potential},
with 10 $\invfb$ at the LHC,
the 5$\sigma$ discovery reach of $W_H$ is possible
only if $\kappa_q>2$ for $f\sim 500$ GeV and extends to $f\sim 650$ GeV for
$\kappa_q\sim 4$.
For smaller values of $\kappa_q$,
3$\sigma$ discovery is possible for $\kappa_q\sim 1$.

On the other hand, suppose we assume $\kappa_{\ell}=0.3$ so that
the cascade $W_{H}\rightarrow \ell L_{-}\rightarrow
\ell\ell^{\prime} A_{H}$ (with one of $\ell$ and $\ell^{\prime}$
being a neutrino) is now allowed.
While the signature of $W_{H}W_{H}$ pair production now still
contains two leptons and missing energy,
the transverse momenta of the leptons are now typically
higher (for the same $M_{W_{H}}$) and the SM background
can be reduced more efficiently with the same cut
on the lepton transverse momentum (see Cao et al.~\cite{Cao:2007pv}
for details).
The results of this search are shown
in the lower two plots of Fig.~\ref{fig:LHT-potential},
which clearly indicates a further reach than the top two plots
(with $\kappa_{\ell}=0.5$) of the same figure.
With $\kappa_{\ell}=0.3$ and 10 fb$^{-1}$ of data,
the LHC can now discover $W_H$
for $f<$ 750 GeV with $\kappa_q=0.5$ at 5$\sigma$ level.
With $\kappa_q=2.0$ and 10 fb$^{-1}$ of data,
the reach for $W_H$ extends to $f<1.1$ TeV.
%

\begin{figure}[h!tp]
\begin{center}
\includegraphics [width=3.15in]{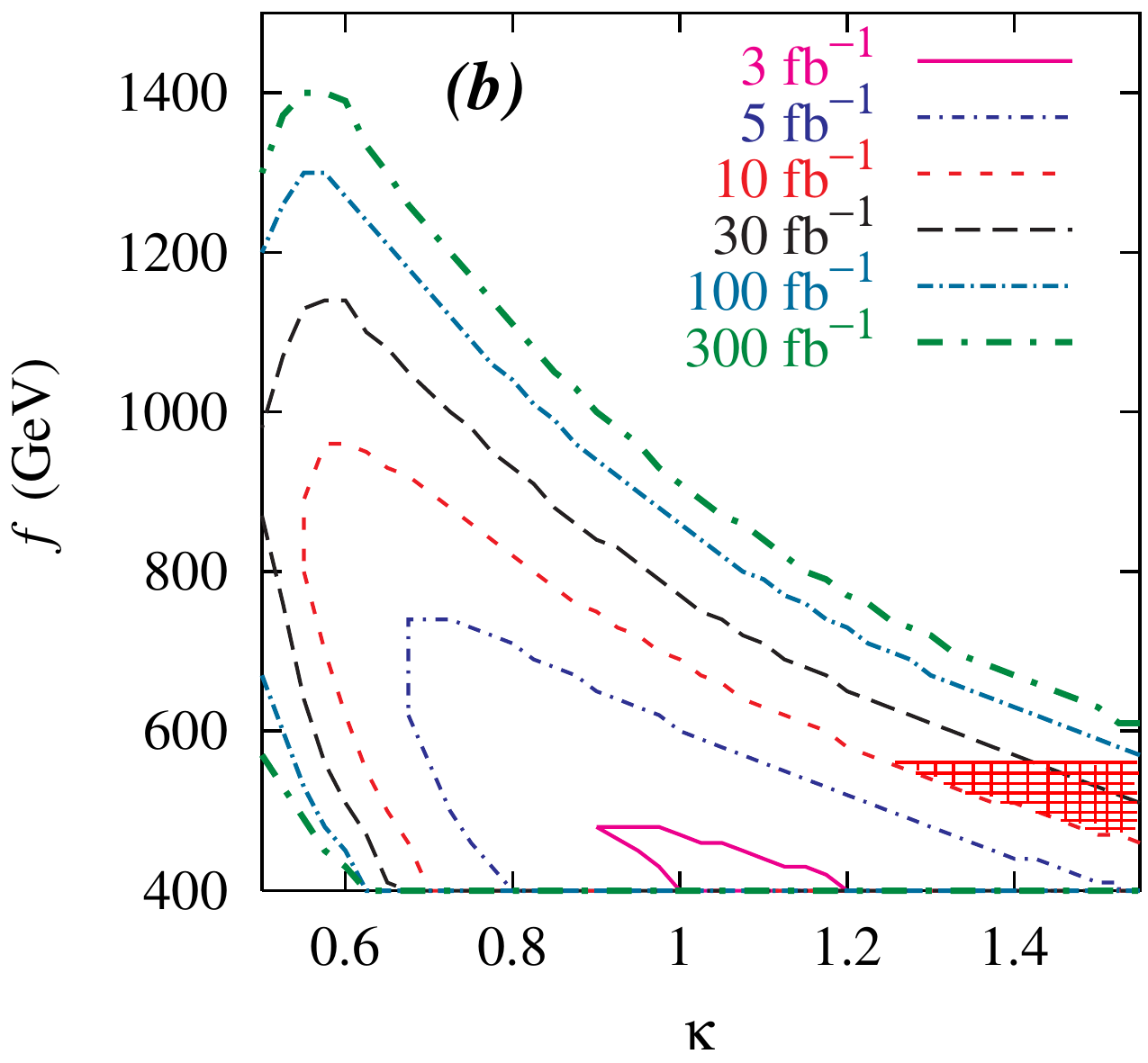}
\caption{
This plot is taken from Choudhury et al.~\cite{Choudhury:2006mp}
and presents the discovery contours of $T$-odd fermions
in on $\kappa$-$f$ plane, where $\kappa_q=\kappa_{\ell}=\kappa$ is
assumed
universal and flavor-diagonal for the first two generations.
We have hashed out (in red) the region of having a viable lone Higgs
scenario with large suppression in $\HGG$ ($f<560$ GeV).}
\label{fig:LHTF}
\end{center}
\end{figure}

\paragraph{$T$-odd quarks and leptons}$\ $\newline
\indent The Tevatron and LHC phenomenology of $T$-odd fermions has been studied in
the literature \cite{Hubisz:2004ft}\cite{Carena:2006jx}
\cite{Choudhury:2006mp}.
The $T$-odd fermions will be pair-produced at the LHC and decay through a
cascade (if kinematically allowed) to $A_H$ or decay to
$A_{H}$ directly, and the missing energy from $A_H$ may
fake the MSSM signatures \cite{Hubisz:2004ft}.
The cascade decay of $Q_{-}$ via
$Q_{-}\rightarrow qW_{H}\rightarrow q\,W\,A_{H}$is possible
as long as $\kappa_q>g(\sqrt{2})^{-1}\sim 0.46$.
Assuming universal and flavor-diagonal $\kappa_q$ for
both the up and down types of quarks, the $5\sigma$
discovery contours of $Q_{-}$ for various
integrated luminosities are
presented on
$\kappa\!-\!f$ plane in Choudhury et al.~\cite{Choudhury:2006mp},
which we show in Fig.~\ref{fig:LHTF} and briefly summarize the results
below.

The cascade decays of pair-produced $\overline{Q}_{-}Q_{-}$
can have the structure
$\overline{Q}_{-}Q_{-}\rightarrow qq W^{+}_{H}W^{-}_{H}
\rightarrow qqW^{+}W^{-}A_{H}W_{H}$, and leptonic decays
of both $W$ gives a signature of
two jets, two opposite-sign leptons, and missing energy.
One can also replace one $W_{H}$ in the cascade by
$Z_{H}$, which decays via $Z_{H}\rightarrow A_{H}(h\rightarrow \overline{b}b)$.
The signature now contains one single lepton, $\overline{b}b$ pair from Higgs decay,
two jets, and missing energy.

In Fig.~\ref{fig:LHTF},
for a given $f$, we can evade the discovery of $Q_{-}$
with a large enough $\kappa$ (heavy enough $Q_{-}$).
Furthermore, the reach of $f$ decreases
as $\kappa$ increases; for example, with 10 fb$^{-1}$, the reach on
$f$ decreases from 950 GeV to 500 GeV as $\kappa$
increases from 0.6 to 1.6.
Beyond $\kappa>1.6$, with 10 fb$^{-1}$, the reach of $f$ is less than 500 GeV,
which is excluded by electroweak precision tests.

It is interesting to note that we need large $\kappa_{q}$
to evade discovery of $Q_{-}$, while we need small $\kappa_{q}$
to evade the discovery of $W_{H}$, and we will explore this further
later in this work.
For now, we simply note that for $f=560$ GeV, $\kappa>1.3$ is required to
evade discovery of $Q_{-}$ at 10 fb$^{-1}$.
%

\paragraph{Top partners}$\ $\newline
\indent In addition to the top quark, the
particle content in the top sector of LHT
includes two top-partners $T_{\pm}$ with
opposite $T$-parity.
The spectrum in the top sector is determined by
the two parameters $\lambda_{1,2}$
in addition
to $f$, with masses given by $m_t\simeq\lambda_1\lambda_2
(\sqrt{\lambda_1^2+\lambda_2^2})^{-1} v_{\smbox{ew}}$,
$m_{T_{+}}\simeq(\sqrt{\lambda_1^2+\lambda_2^2})f$,
and $m_{T_{-}}\simeq\lambda_2 f$,
so that $T_{-}$ is always lighter than $T_{+}$.
The collider signatures of $T_{-}$ have
seen investigated in by Matsumoto et al.~\cite{Matsumoto:2006ws},
the LHC reach of the $T$-odd top quark is about 900 GeV with 50 $\invfb$
\cite{Matsumoto:2006ws}.
On the other hand, this does not translate to
a bound on $f$ that we seek for since one can
always make both top-partners heavy by making
$\lambda_2$ large, and adjust $\lambda_1$
to accommodate the top quark mass (at
the expense of fine-tuning).

\paragraph{Summary}$\ $\newline
\indent To find out whether
LHT with large deviation in $\HGG$
($f<560$ GeV and $m_h\sim 130$ GeV) is consistent
with a lone Higgs scenario,
we combine the results of the previous subsections.
We note from Fig.~\ref{fig:LHTF} that
the discovery potential of the $T$-odd fermions
with 10 $\invfb$ limits a universal $\kappa$
to be either $\kappa < 0.55$ or $\kappa > 1.3$.
We discard the region $\kappa < 0.55$ with
$f\sim 500$ GeV
from our considerations for lone Higgs scenario
because this region contains a rather light $Q_{-}$
with a mass of about 400 GeV, which was shown to be
difficult to detect at the LHC.
With such light $Q_{-}$, although the process
considered
in Choudhury et al.~\cite{Choudhury:2006mp}
($pp\rightarrow Q_{-}\overline{Q}_{-}\rightarrow
q\overline{q}W^{\pm}_{H}Z_{H}\rightarrow
bb+jj+\ell^{\pm}+\displaystyle{\not}E_T$)
can not yield results
that significant enough to claim discovery,
we suspect other optimized searches dedicated to this
region of parameter space may discover $Q_{-}$,
and this particular region
of parameter space may warrant further study.

We start our search for a lone Higgs
scenario with $\kappa_{\ell}=0.5$,
so that $W_{-}$ only decays to
$W_{-}\rightarrow WA_{-}$, and the viable
region in the $\kappa_q-f$ plane is shown in Fig.~\ref{fig:LHT-potential}(a).
On the other hand, discovery reach
for the $T$-odd quarks indicates the viable region as shown in Fig.~\ref{fig:LHTF}.
For $f=560$ GeV, the lone Higgs scenario
constraints $1.25<\kappa_{q}<3$
from the 5$\sigma$ discovery potentials
of $W_{-}$ ($\kappa_q<3$) and $Q_{-}$ ($1.25<\kappa_q$).
Thus, there is a consistent lone Higgs scenario
with large deviations in $\HGG$ in LHT,
and we will discuss the phenomenology of
the lone Higgs scenario in the next section.


\subsection{Minimal Universal Extra Dimension}
\subsubsection{$\HGG$ in MUED}
\label{sec:MUED-HGG}
The Universal Extra Dimension (UED) model \cite{Appelquist:2000nn}
extends the spacetime
with one additional spatial, flat dimension that is accessible to all
fields of the SM (hence the name universal)
\cite{Antoniadis:1990ew}\cite{Antoniadis:1998ig}\cite{ArkaniHamed:1998rs}\cite{ArkaniHamed:1998nn}.
This extra dimension is
compactified on a circle with radius $R$ and orbifolded with a $Z_2$ symmetry.  The SM particles
are zeroth Kaluza-Klein (KK) modes, and the higher KK modes have tree-level masses
roughly $n R^{-1}$, where $n$ is the KK number.
However, the masses of the KK modes are renormalized by interactions
localized at the orbifold fixed points \cite{Georgi:2000ks}.
These effects are scale-dependent and are thus generated by
renormalization effects.
In Minimal UED (MUED) \cite{Cheng:2002iz},
an ansatz is made about the values of
these boundary interactions at the cutoff scale,
and the model is parameterized by two
free parameters: $R$ and the cutoff scale $\Lambda$.

In MUED, a KK-parity is conserved such that the lightest
KK-odd particle (LKP) is stable and can
serve as dark matter.
In the particular case of MUED, the lightest KK-odd particle
is a mixture of (dominantly) the first KK modes of
the hypercharge gauge boson $B^{(1)}_{\mu}$
and (sub-dominantly) the neutral $SU(2)$ gauge boson
$W^{(1)}_{3\mu}$
\cite{Cheng:2002ej}\cite{Servant:2002aq}\cite{Servant:2002hb}\cite{Kong:2005hn}\cite{Burnell:2005hm}.
The Wilkinson Microwave Anisortopy Probe (WMAP) observations \cite{Spergel:2006hy} of the dark matter
relic density translates into a tight
constraint on $R^{-1}$ of $500\mbox{\ GeV}< R^{-1}< 600\mbox{\ GeV}$,
and this range of $R^{-1}$ in MUED implies
colored KK modes with sub-TeV masses,
so that they are accessible at the LHC.

Some typical spectra of MUED can be found in Cheng et al.~\cite{Cheng:2002iz},
and a review of the collider signatures of generic UED models is presented by
Hooper et al.~\cite{Hooper:2007qk}.
Here we simply reproduce formula for the masses of the KK particles
relevant to our discussion
with the MUED ansatz in Cheng et al.~\cite{Cheng:2002iz}.
We are interested in the mass of the KK $W^{\pm}$-boson and
the two KK top quarks (corresponding to KK modes of the
left- and right-handed components of the SM top quark)
because they contribute to the di-photon
decay of the Higgs boson, and the KK top quarks also contribute
to the gluon-fusion production of the Higgs boson,
We are also interested in the mass of the KK gluon as we will
use it to analyze the LHC discovery reach on the parameter $R^{-1}$.
The masses of the $n^{\smbox{th}}$ KK gluon and KK $W^{\pm}$-boson are given by
\begin{align}
m^2_{g^{(n)}}&=n^2 R^{-2}\left(1+\frac{23}{2}\frac{g_3^2}{16\pi^2}\ln\frac{\Lambda^2}{\mu^2}\right),\\
m^2_{W^{(n)}}&=n^2 R^{-2}\left(1+\frac{15}{2}\frac{g^2}{16\pi^2}\ln\frac{\Lambda^2}{\mu^2}\right),
\label{eq:MUED-KKG}
\end{align}
where $g_3$ is the strong coupling.
At every non-zero KK level there are
two KK top quarks with the mass matrix
\begin{align}
\begin{pmatrix}
n R^{-1} + \delta m_{Q_3^{(n)}} & m_t \\
m_t  & -n R^{-1}- \delta m_{U_3^{(n)}}
\end{pmatrix},
\label{eq:MUED-KKTOP}
\end{align}
where the radiative corrections are given by
\begin{align}
\delta m_{Q_3^{(n)}} &= n R^{-1}
\left(
3\frac{g_3^2}{16\pi^2}
+\frac{27}{16}\frac{g^2}{16\pi^2}
+\frac{1}{16}\frac{g^{\prime 2}}{16\pi^2}
-\frac{3}{4}\frac{y_t^2}{16\pi^2}
\right)
\ln\frac{\Lambda^2}{\mu^2},\\
\delta m_{U_3^{(n)}} &= n R^{-1}
\left(
3\frac{g_3^2}{16\pi^2}
+\frac{g^{\prime 2}}{16\pi^2}
-\frac{3}{2}\frac{y_t^2}{16\pi^2}
\right)
\ln\frac{\Lambda^2}{\mu^2},
\end{align}
where $g$ and $g^{\prime}$ are the electroweak
gauge couplings.
We choose the renormalization scale to be $\mu=R^{-1}$ and
$\Lambda= 10R^{-1}$, so the masses of the first KK modes
of the gluon and $W$-boson are approximately
\begin{align}
m_{g^{(1)}}&\simeq 1.23\, R^{-1},\\
m_{W^{(1)}}&\simeq 1.05\, R^{-1}.
\label{eq:MUED-eq-simp1}
\end{align}
For the first KK modes of the top quark, having $\mu=R^{-1}$
and $\Lambda= 10R^{-1}$ gives us the mass matrix
\begin{align}
\begin{pmatrix}
1.13\, R^{-1} & m_t \\
m_t & -1.09\, R^{-1}
\end{pmatrix},
\end{align}
and when $R^{-1}\gg m_t$, the mass eigenvalues can
be approximated by the diagonal entries
\begin{align}
m_{t_2^{(1)}}&\simeq 1.13\, R^{-1},\\
m_{t_1^{(1)}}&\simeq 1.09\, R^{-1}.
\label{eq:MUED-eq-simp2}
\end{align}

In MUED, the strengths of the gauge and Yukawa interactions of the Higgs boson
are the same as those in the SM.
However, its effective di-gluon and di-photon
couplings differ significantly from those in the SM because of the
additional contributions induced by the
KK partners of the top quark and the $W$-bosons.
The effects of these KK modes have been computed by Petriello
\cite{Petriello:2002uu},
and it is found that the presence of KK top quarks always enhances
the production rate of Higgs boson via gluon-gluon fusion.
We adapt the results of this
reference to \verb"hdecay" \cite{Djouadi:1997yw},
taking into account the radiative corrections to
the masses of the KK top quarks.
(Including
the radiative corrections, however, does not qualitatively modify
the conclusions of Petriello \cite{Petriello:2002uu}.)
%

\begin{figure}[h!t]
\begin{center}
\includegraphics [width=3.15in]{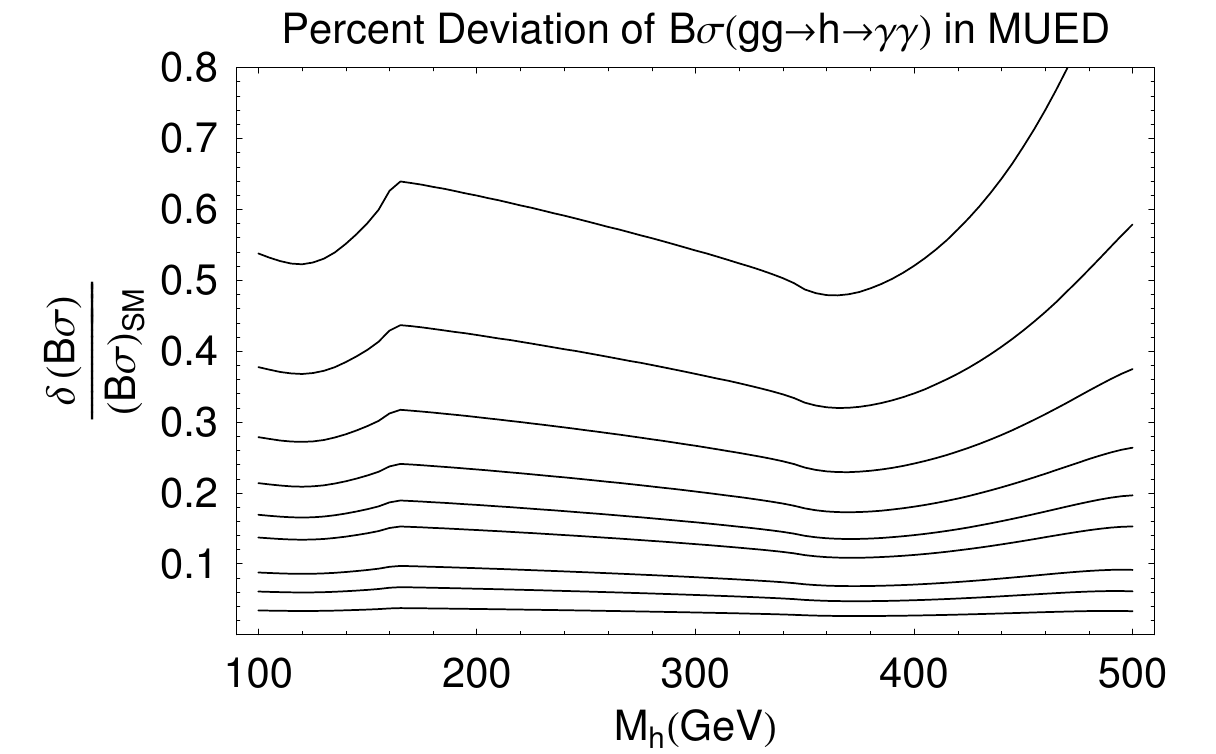}
\includegraphics [width=3.15in]{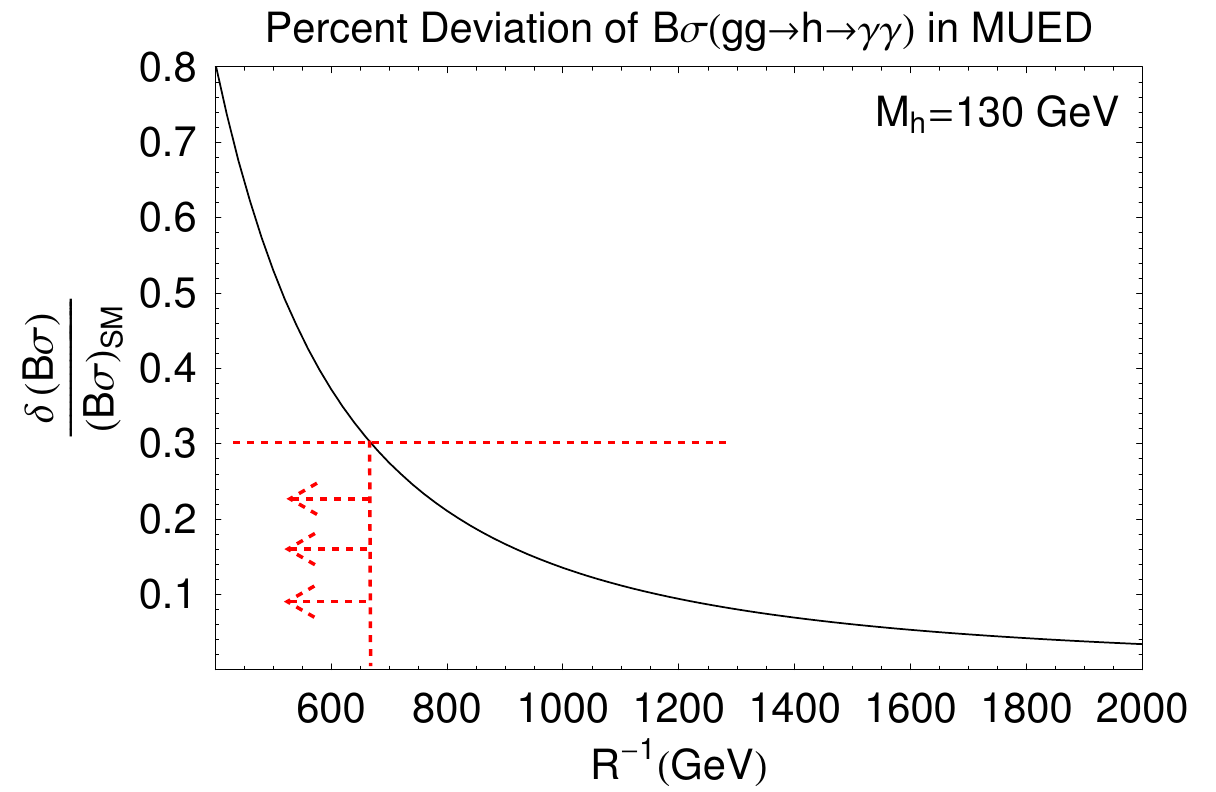}
\caption{The plot on the left shows the fractional deviation of $\HGG$ as a function
of Higgs mass in MUED for various values of $R^{-1}$.
From top to bottom, the values of $R^{-1}$ (in GeV) are respectively
500, 600, 700, 800, 900, 1000, 1250, 1500, and 2000.
In each case,
we choose $\Lambda = 10 R^{-1}$, and sum over
the contributions from the lowest 10 KK levels.
The plot on the right shows the fractional deviation of $\HGG$ as a function
of $R^{-1}$ in MUED for $m_{h}=130$ GeV, and only for
$R^{-1}< 650$ GeV is there a large enhancement in $\HGG$.
}
\label{fig:MUEDPlot}
\end{center}
\end{figure}
%

In the left plot of Fig.~\ref{fig:MUEDPlot},
we show the deviation signal $\HGG$ from the SM
values for several values of $R^{-1}$ as a function
of the Higgs mass.
We first note that
for $500\ \mbox{GeV}<R^{-1}<600\ \mbox{GeV}$ the product $\HGG$
is at least 35\% (and at most 70\%) above the SM results for
Higgs mass in the range of
$100\ \mbox{GeV}<m_h<200\ \mbox{GeV}$.
Together with the accessibility of the KK top quarks and KK gluons
(with masses on the order of 700 GeV) at the LHC,
this scenario can be distinguished from the SM.
Furthermore, for $R^{-1}>$ 700 GeV, the signal enhancement
is less than 30\%, and it would be difficult to distinguish
the signal from the SM.  In this case, the UED model may still
explain dark matter with mass spectra that deviate from
the MUED ansatz, and $R^{-1}$ in Fig.~\ref{fig:MUEDPlot}
should be interpreted as the mass scale
of the KK top quarks and gauge bosons through
Eqs.~(\ref{eq:MUED-eq-simp1}) and (\ref{eq:MUED-eq-simp2}).

Since the di-photon branching ratio is significant
only if the Higgs boson is significantly lighter than the $WW$
threshold, we focus on a light Higgs boson with
$m_h\lesssim 130$ GeV.
From Fig.~\ref{fig:MUEDPlot}, we also see that the
deviation of the signal is roughly independent of Higgs mass
in the range of 100 GeV$< m_h <$ 140 GeV, and
we plot the deviation in signal $\HGG$ from the SM
values as a function of $R^{-1}$ in the second plot of
Fig.~\ref{fig:MUEDPlot} for $m_h=130$ GeV,
where we also see that
for $R^{-1}>$ 700 GeV, the deviation of the signal is less
than 30\%.
Thus, the viable region in the MUED model
for producing sizeable ($>30\%$) enhancement
in $\HGG$ as compared to SM
is $R^{-1}\lesssim$ 700
GeV, and we turn to a discussion about the discovery reach of
the various KK resonances at the LHC to see
if this region of $R^{-1}$ can be consistent with a lone
Higgs scenario.

\subsubsection{Lone Higgs Scenario in the MUED}
\label{sec:MUED-LH}
Qualitatively, the particle content and interactions of the MUED model
is similar to MSSM in many ways in that each
particle in the SM is extended with a partner.
Instead of superpartners
with spins that differ by half integer in MSSM, the MUED
has KK modes with the same spin.  The role of $R$-parity
in the MSSM is played by $K$-parity in the MUED.
In addition to giving rise to a dark matter candidate,
such parity ensures that these partners must be
pair-produced at the LHC.

Assuming that the colored KK states are heavier
than the non-colored KK states (as is the case in the
MUED ansatz), we would expect the LHC collider
signatures of the KK gluons and quarks to follow
similar paths as those of the gluino and squarks of the MSSM.
The KK gluons/quarks would be pair-produced and decay
through a cascade that ends with the LKP.
As with the case of the MSSM, the signature would again be jets with missing energy,
and we can reasonably approximate the discovery
reach of the KK gluon and quarks to be the same as
the gluino and the squark of the MSSM,
which is 2.4 TeV for the mass of the gluino.
For discovery reach of the KK quarks, we make the same simplifying
assumption that we made for the discover reach of the squarks of the
MSSM and assume that the reach for the KK quarks is also
2.4 TeV.

Since $R^{-1}\sim$ 700 GeV corresponds to a KK gluon with
a mass of approximately 860 GeV (see Eq.~(\ref{eq:MUED-eq-simp1})),
the KK gluon is within the reach
of LHC using MSSM gluino discovery potential as a guide.
We therefore do not have a consistent lone Higgs
scenario with $R<700$ GeV.
To see what types of lone Higgs scenarios
can be consistent with the MUED spectra,
we note that once $R^{-1}$ is large enough so that
we enter the lone Higgs scenario ($M^{(1)}_g \gtrsim 2.4$ TeV
so that $R^{-1}\gtrsim$ 1.95 TeV via Eq.~(\ref{eq:MUED-eq-simp1})),
the signal $\HGG$ is enhanced only by 5\% (see Fig.~\ref{fig:MUEDPlot}),
which is well below
the expected sensitivity of the LHC, even with 100 $\invfb$ of data.
This result should be contrasted with MSSM,
where decoupling the s-top and gauginos
still allowed a large suppression in $\HGG$.
As stated earlier, in the MSSM this suppression
comes from $\tan\beta$-enhanced decay widths
$\Gamma(h\rightarrow \overline{b}b)$
and
$\Gamma(h\rightarrow \overline{\tau}\tau)$,
and thus a suppression in the branching ratio
$\mbox{Br}(h\rightarrow\gamma\gamma)$.
In MUED, the couplings of the Higgs
boson to the SM fermions are the same as those
in the SM, and the deviation in $\HGG$ comes
only from the KK top quarks and gauge bosons contributions,
which decouples accordingly as $R^{-1}\rightarrow\infty$.

The MUED ansatz with $R^{-1}\gtrsim$ 1.95 TeV also
means that the KK leptons and gauge bosons
are out of the reach at the LHC, since the masses
of all these particles are of the order $R^{-1}$.
Once we move away from the
MUED ansatz, however, the KK states can in principle have
independent masses and the
discovery reach of the KK states
do not translate to bounds on $R^{-1}$ without
a more fundamental organizing principle.
Since in this work we study the signal $\HGG$
only using the MUED ansatz with $\Lambda$ fixed
as $\Lambda=10R^{-1}$, without worrying about
constraints and implications of the LKP
as dark matter, we will simply report that a lone Higgs scenario
is consistent with the MUED ansatz with $R^{-1}>1.95$ TeV,
with the reasonable assumption that
the discovery reach of the KK gluon is the same as
the gluino of the MSSM.
However, in this case, the MUED new physics signal
can not be easily distinguished from the SM.
We will discuss how to distinguish the various models
of new physics in the next section.

\section{Distinguishing the Models}
\label{sec:HGG-lone}

\subsection{Brief summary of results}
\begin{table}[h!t]
\begin{center}
\caption{The possibilities
of new physics model based on the deviation
of $\HGG$ from the SM, without imposing constraints
of being consistent with a lone Higgs scenario.}
\label{tb:general}
\vspace{0.125in}
\begin{tabular}{|l|l|l|}
\hline
  &  $m_h <$130 GeV   & $m_h >$130 GeV \\ \hline
Enhanced   & MUED         & MUED  \\ \hline
Similar    & MSSM, MUED, LHT & LHT, MUED\\ \hline
Suppressed & MSSM, LHT         & LHT \\ \hline
\end{tabular}
\end{center}
\end{table}

At this point, we summarize our findings so far.
As noted earlier, it is useful to focus in the range of $m_h <$130 GeV
because the lightest CP-even Higgs can not be heavier
than this bound in the MSSM, and, above this mass range the di-photon
branching ratio drops, and it will be more useful to examine other
modes of decay.
First, we do not restrict ourselves to the lone
Higgs scenario and include the possibilities
that new physics is light and there can be significant
deviations of the signal.
This is presented in Table \ref{tb:general}.
In addition, from our analysis earlier, we also
expect to see additional resonances when there
is a significant deviation of $\HGG$ compared to
the SM.

\begin{table}[h!t]
\begin{center}
\caption{
The possibilities
of new physics model based on the deviation
of $\HGG$ from the SM,
imposing the conditions
of being consistent with the lone Higgs
scenario, i.e.~that no new physics is seen at the LHC
with 10 fb$^{-1}$ of integrated luminosity and $m_h<130$ GeV.}
\label{tb:first-step}
\vspace{0.125in}
\begin{tabular}{|l|l|}
\hline
  &  $m_h <$130 GeV   \\ \hline
Enhanced   & -        \\ \hline
Similar    & MSSM, MUED, LHT \\ \hline
Suppressed & MSSM, LHT \\ \hline
\end{tabular}
\end{center}
\end{table}

Stepping towards the lone Higgs scenario,
we impose the conditions that the Higgs is lighter
than 130 GeV and no additional new resonances
has been directly seen at the LHC after 10 $\invfb$
of operation.
This is presented in Table \ref{tb:first-step}.
As noted earlier, decoupling the
KK tops and gauge bosons in MUED such that they are
not directly accessible at the LHC after 10 $\invfb$
necessarily implies that the resulting Higgs
boson would have a signal that is similar to the SM.
In LHT with $m_h<130$ GeV, the deviation of $\HGG$ is only more
than 30\% in the limited ranges of $f <$ 560 GeV and $m_h\sim 130$
GeV, and this region of parameter space leads to a viable lone
Higgs scenario only for $1.3<\kappa_{q}<3.0$.
The MSSM can also lead to a viable lone Higgs
scenario with large deviation in $\HGG$, for example,
with $M_A=300$ GeV and $\tan\beta=10$.

Although we noted earlier that,
relative to the SM, the signal $\HGG$ is always
enhanced in MUED and suppressed in LHT,
in a lone Higgs scenario we would not
expect MUED to show a significant enhancement.
If we see a suppression in the signal,
we have to distinguish between LHT and the MSSM.
Before we attempt to distinguish between the LHT
and the MSSM based on large deviations in
$\HGG$ alone, we first address
what would happen if the LHC observes
a Higgs boson with a deviation in $\HGG$
that is less than 30\%.

\subsection{Lone Higgs Scenario with a small deviation in $\HGG$}
\label{sec:LH-small}
Suppose that the LHC finds a Higgs boson after 10 $\invfb$
whose $\HGG$ measurement deviates less than 30\% from
the expected SM value.
Though we can not directly distinguish
between the three new models of
physics from the SM,
our results may nonetheless be useful in
devising further tests to distinguish the models, for example,
with consistency checks.
Broadly speaking, if the deviation
of the signal is more than the eventual
accuracy of 10\%, then it may be worthwhile
to pursue this deviation as both a lead for
consistency checks and a bias for planning
search strategies/signals that are more optimized
to one particular model of new physics.
That is, we are placing more
confidence in the measurement
than its uncertainty warrants,
in the hope that subsequent measurements
would improve the uncertainty without a great
shift in the measured value.

For example, suppose that with 10 $\invfb$ of
LHC data,
the measured
value of $\HGG$ shows a positive 20\% deviation from
the SM with a 30\% uncertainty.
While this is not a significant deviation
from the SM and this deviation is likely to
fluctuate (and may even change its sign)
with subsequent measurements,
we can use this result to favor MUED
over LHT and the MSSM, since neither of these
models can give an enhancement in $\HGG$.
As such a deviation corresponds
to $R^{-1}$ in the range of 800 to 900 GeV and
the masses of the KK tops and KK gluons of about 1 TeV,
we may use this deviation as a consistency check to
rule out the MUED model if none
of the heavy KK particle states is found,
and seek other explanations
for the enhancement in the Higgs signal.

As another example, suppose that we see a 20\%
suppression of the signal from the SM with
10 $\invfb$ of LHC data, we can then
favor LHT and the MSSM over MUED,
and optimize our search strategies in order
to both find additional expected resonances in these
models as well as devise tests that may distinguish
these two models when we find these resonances.
To be more concrete, let us also suppose that the Higgs
mass is measured to be 125 GeV.
Such Higgs mass is rather large in the scope of the MSSM,
and is consistent with
heavy s-tops, with perhaps significant
mixing that evade direct discovery.
If we favor the assumption
Nature is supersymmetric,
from Fig.~\ref{fig:MSSMTPlot}, we see that a
suppression of 20\% in $\HGG$ corresponds to $M_A\sim 400$ GeV.
On the other hand, in the context of LHT,
from the Fig.~\ref{fig:LHTPlot},
we see that such deviation correspond to a rather low
value of $f\sim 640$ GeV.
These pieces of information ($M_A\sim 400$ GeV
or $f\sim 640$ GeV) can then be used to devise
search strategies that can differentiate MSSM from
LHT.

\subsection{Lone Higgs Scenario with a large deviation in $\HGG$}
\label{sec:LH-big}
If the LHC finds a lone Higgs boson with
$\HGG$ that deviates from the SM value
by more than 30\%, then the prospects are very exciting.
First, in the case of an enhancement in the measurement,
we can strongly disfavor all three models of new physics
under considerations here, and seek alternative
explanations for such an enhancement.
Neither the MSSM nor the LHT can give a enhancement
in $\HGG$ relative to the SM in a lone Higgs scenario,
and this large enhancement can not be consistent
with MUED because in that case some light KK
resonances must be produced and detected at the LHC.
(By definition, lone Higgs scenario requires that
no other resonances being detected.)

In the case of a suppression,
we would strongly favor MSSM and LHT over MUED
and need to distinguish these two models
based on the measured value of $\HGG$.
As a consistency check with both MSSM
and LHT, the Higgs mass should be
around $m_h\sim 125$ GeV to be consistent with
both the MSSM (because the s-tops need to be heavy
to evade discovery) and LHT (because of the large
deviation in $\HGG$).

A possible first discriminating test between MSSM and LHT
is the amount of deviation in $\HGG$.
From Fig.~\ref{fig:LHTPlot}, the LHT with $m_h\leq 130$ GeV can not
give a suppression in $\HGG$ of more than 40\%, while
the MSSM with $10\leq\tan\beta\leq 30$ can accommodate
a suppression larger than 40\% with
$M_A\lesssim 300$ GeV as shown in Fig.~\ref{fig:MSSMTPlot}.
Therefore, a lone Higgs scenario with a suppression
in $\HGG$ of more than 40\% as compared to the SM
can only be consistent with the MSSM.
(As discussed in Section \ref{sec:MSSM-HGG},
if we impose minimal flavor violation,
the 3$\sigma$ uncertainties in the
measurement of $\mbox{Br}(b\rightarrow s\gamma)$ imposes
$M_A>260$ GeV, and the suppression in $\HGG$
can not be more than 55\% relative to the SM.)
While the suppression between
30\% and 40\% in $\HGG$ relative to the SM prediction
can be consistent with both the MSSM and the LHT,
we propose possible tests that can distinguish
these two models with more integrated luminosity in the next subsection.

\subsection{Distinguishing different lone Higgs scenarios with more luminosity}
\label{sec:MSSM-vs-LHT}

Given a lone Higgs scenario with a measured suppression in $\HGG$ that
is greater than 30\% relative to the SM,
we note in the previous subsection that
\begin{itemize}
\item
if the suppression is greater than 40\% relative to the SM, then
MSSM is strongly favored, and,
\item
if the suppression in $\HGG$ relative to the SM is
between 30\% and 40\%, then both the MSSM and the LHT are viable.
\end{itemize}
In this subsection, we discuss how
we may distinguish the MSSM from the LHT when
the suppression in $\HGG$ relative to the SM is
between 30\% and 40\%
with further LHC operations and seeking new resonances
beyond 10 fb$^{-1}$

First we consider the case that the LHT is responsible for
the observed suppression in $\HGG$ relative to
the SM.
In this case, a suppression of
30\% or more in $\HGG$ relative to the SM prediction
corresponds to $f< 560$ GeV.
For $f=560$ GeV,
the viable range of $\kappa_q$ is
$1.25 < \kappa_q < 3$, and
this viable range of $\kappa_q$ becomes even smaller with smaller $f$.
For example, with $f=500$ GeV,
the viable range of $\kappa$ is
$1.6 < \kappa_q < 2.2$.
With more luminosity,
the LHC should discover
either $Q_{-}$ and/or $W_H$ because
this viable range on $\kappa$ disappears,
and we no longer have a lone Higgs scenario.
For example, with $f=560$ GeV and
100 fb$^{-1}$ of data, the
LHC can discover $W_H$ if $\kappa_q>1.5$
(see Fig.~\ref{fig:LHT-potential}(b))
and discover $Q_{-}$ if $\kappa_q<1.55$
(see Fig.~\ref{fig:LHTF}).
These two constraints of $\kappa_q$ overlap,
so at least one of $W_H$ or $Q_{-}$ should
be discovered at 100 fb$^{-1}$,
and if $\kappa$ is between $1.5<\kappa_q<1.55$,
then the LHC should discover both
$W_{H}$ and $Q_{-}$ with 100 fb$^{-1}$.

If we assume that the MSSM is the model of new physics
that underlies the observed lone Higgs scenario with
a suppression in $\HGG$ between 30\% and 40\%,
then Fig.~\ref{fig:MSSMTPlot} tells us that
$290\,\mbox{GeV}\lesssim M_A\lesssim 350\,\mbox{GeV}$, and
Fig.~\ref{fig:NeutralHiggs}(c) gives
us the viable region on $M_A-\tan\beta$ plane for a lone
Higgs scenario.
From Fig.~\ref{fig:NeutralHiggs}(c), we also
see, in some of the viable region, with higher luminosity
the LHC has the potential to discover the heavy
MSSM Higgs bosons $H/A$ in the $H/A\rightarrow \overline{\tau}\tau$
mode.
Unfortunately, unlike the case of the LHT, we are not
guaranteed to discover additional resonances at the LHC
with more luminosity.
For example, with $\tan\beta=10$, $M_A=300$ GeV, and
10 fb$^{-1}$ of LHC data,
we have a lone Higgs scenario with a significant suppression
in $\HGG$ compared to the SM prediction,
but we are outside the LHC reach for the heavy MSSM bosons
$H/A$ even at 300 fb$^{-1}$ of integrated luminosity.

\begin{figure}[h!t]
\begin{center}
\includegraphics [width=3.0in]{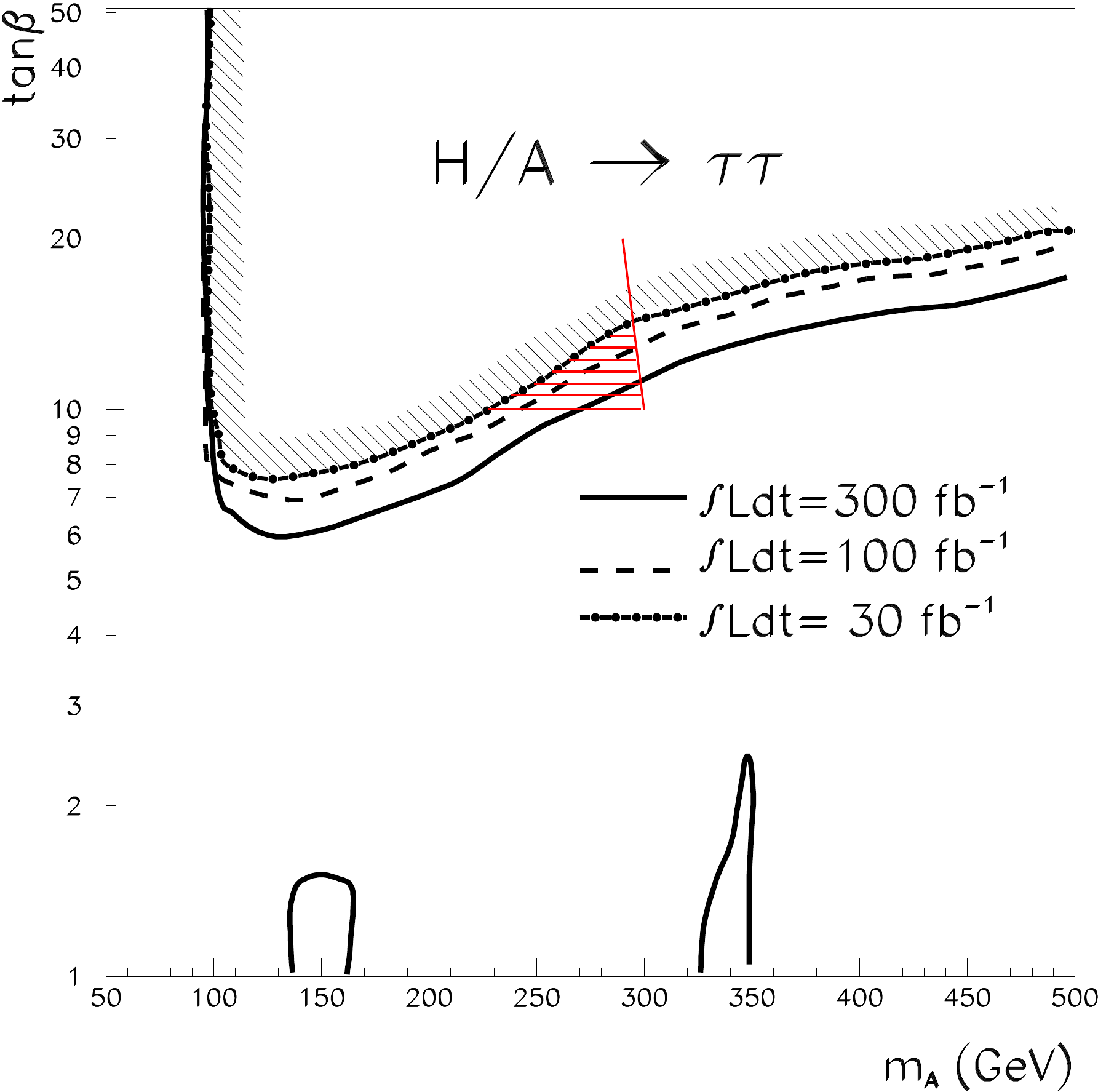}
\caption{We superimpose the discovery reach
of $H/A$ in the ATLAS TDR (Fig.~\ref{fig:NeutralHiggs}(c))
with the upper bounds on $M_A$ from a suppression of 40\%
in $\HGG$ (the negatively-sloped line).
The hashed region below the intersection
is the viable lone Higgs scenario region
with a suppression in $\HGG$ of at least
40\%.  We assume $\tan\beta>10$ throughout
our analysis.
}
\label{fig:LHS-MSSM-BIGDEV}
\end{center}
\end{figure}

We can also use the discovery of the heavy Higgs bosons $H/A$
as a consistency check of MSSM for a lone Higgs scenario with
a suppression in $\HGG$ of more than 40\% relative to the SM.
Suppose we have a lone Higgs scenario with
a measurement of $\HGG$ that is suppressed by more than 40\% relative to the SM.
From Fig.~\ref{fig:MSSMTPlot}, we see that
such suppression corresponds to $M_A\lesssim 300 (290)$ GeV
if $\tan\beta=10(20)$.
In Fig.~\ref{fig:LHS-MSSM-BIGDEV},
we superimpose onto Fig.~\ref{fig:NeutralHiggs} with
the contour that corresponds to having a 40\% suppression in
$\HGG$ relative to the SM and mark the viable
region of parameter space.
We see that in most of this parameter space, we have the potential
to discover the heavy MSSM neutral Higgs boson $H/A$ when
the integrated luminosity is increased to 100 fb$^{-1}$
if $\tan\beta$ is sufficiently large.
And with 300 fb$^{-1}$ of integrated luminosity, most of this
parameter space is covered.
%
%

\subsection{Other parameter-independent tests to distinguish the models}
\label{sec:general-test}
So far, we have attempted to use
the lone Higgs scenario
and a large deviation in the measurement
of $\HGG$ to discriminate between
MUED, LHT, and the MSSM.
For some states in these models of new physics, such as the
s-leptons of the MSSM, we can obtain viable lone Higgs scenarios
by simply raising the masses of the new states large enough to
evade discovery without affecting significantly the phenomenology
of the discovered boson.
It is possible that
these states are discovered and identified
only after 10 $\invfb$ of integrated luminosity,
and these situations can certainty help us to further
discriminate
the models of new physics.

In this section we give some more tests of
this type that can discriminate between MUED, LHT, and the MSSM,
independent of the lone Higgs scenario.
(Some of these tests have been stated in the literature.)

Since the MSSM is the most extensively studied
model of new physics, and it has arguably
the most complicated collider phenomenology of the
three models that we study in this work,
we present tests that may separate
LHT and MUED from the MSSM rather than identify
MSSM directly.

\subsubsection{Distinguishing MSSM and LHT with single-$T_{+}$ Production}
In the MSSM, the cancellation of quadratic
divergence introduces new states
that differ in spin and must be pair-produced
due to $R$-parity.
In LHT, the new physics that modifies the properties
of the Higgs boson come from the top-partners
($T$-parity even or odd) and the $T$-odd
gauge bosons ($W^{\pm}_H$).
The discovery potentials of the $T$-odd gauge
bosons and top-partners were summarized
in Section~\ref{sec:LHT-LH},
and it was pointed out that, while
the masses of the top-partners depend on
additional inputs $\lambda_{1,2}$
(with the top quark mass as one constraint),
the top-partners' contributions to $\HGG$
is independent of these parameters.
Although we could raise
the masses of both top-partners
to evade LHC discovery at 10 $\invfb$,
naturalness arguments favor light top-partners
and it is not implausible that we
discover the top-partners after 10 $\invfb$
of operation.
In naturalness arguments, the role of $T_{+}$
in LHT is played by the s-tops $\tilde{t}_{1,2}$
in the MSSM to cancel the quadratic divergences
in the Higgs boson self energy, and, assuming we see new resonances
related to an extended top-sector, we would like to distinguish $T_{+}$
and $\tilde{t}_{1,2}$.

One potential signature that can be used
to distinguish LHT model from both the MSSM and MUED
is the single-production of the $T$-even top-partner ($T_{+}$) \cite{Han:2003wu}.
The production mechanism of such signature
is similar to the SM single-top production, with
the top quark replaced by $T_{+}$.
As new colored states in both the MSSM
and MUED must be pair produced,
this can be used to distinguish the
LHT from the MSSM and MUED.
Furthermore, in the LHT the
single-top production rate is suppressed as compared
to the SM rate,
and a precise measurement
in single-top production rate can give
us information about the top sector
in LHT \cite{Cao:2006wk}.

\subsubsection{Distinguishing MSSM and MUED with single-$H/A$ production}
As stated earlier, the additional particle content in the MUED is
very similar to that of the MSSM: the KK-modes are partners
to the SM particle content in MUED just as the super-partners
in the MSSM.
Where as the gauginos obtain SUSY-breaking
mass in the MSSM, in MUED models, the $n^{\smbox{th}}$ KK
mode of the gauge bosons ($A_{\mu}^{(n)}$)
become massive through a `geometric' Higgs mechanism:
they eat linear combinations
of (dominantly) the $n^{\smbox{th}}$
fifth-dimensional component of the gauge bosons ($A_5^{(n)}$)
and (sub-dominantly) the $n^{\smbox{th}}$ mode of the Higgs
boson $H^{(n)}$.
At the first KK level, there are then
four physical scalar bosons from the un-eaten
combinations that are dominantly the first
KK mode of the Higgs boson: $H^{\pm (1)}$,
$\mbox{Re}[H^{0(1)}]$, and
$\mbox{Im}[H^{0(1)}]$.
Furthermore, in the limit that $R^{-1}\gg M_Z$
and if we assume no significant radiative corrections
to the Higgs masses,
these Higgs bosons will have similar masses with
fractional
degeneracies of the order $\mathcal{O}(m_h^2/R^{-2})$.

Effectively, the MUED also contains a two Higgs doublets
just as the MSSM: the zeroth and first KK-level Higgs doublets.
The Higgs sector in the first KK level of MUED models
is similar to the MSSM heavy Higgs sector
in many aspects:
\begin{itemize}
\item they both contain two electrically neutral (one $CP$-even
and the other $CP$-odd) Higgs bosons, and a $Q_{em}=1$ charged Higgs boson,
\item these states are
nearly degenerate: the fractional degeneracies in the masses are of the order $
\mathcal{O}(M_Z^2/M_A^2)$
in the MSSM, and $\mathcal{O}(m_h^2/R^{-2})$ in MUED.
\end{itemize}
However, there is one crucial difference:
the heavy Higgs bosons in MSSM are
$R$-even and can be singly produced at the LHC,
whereas these KK Higgs are
$K$-odd and must be pair produced.

Unfortunately, as the LHC discovery potential
for the singly-produced heavy
Higgs bosons in the MSSM do not place stringent
constraints on parameter space (this allows
us to have a lone Higgs scenario with
a large deviation in $\HGG$),
discovering new physics through
pair-produced Higgs bosons
is much more difficult.
Nevertheless, we point out this crucial
difference in the Higgs sector
in the MSSM and MUED since, as far as we know, it had
not been explicitly stated in the
literature.
Furthermore, it could generate
new signal event signatures outside the context of
lone Higgs scenario.

\section{Conclusions}
\label{sec:conc}
In this work we attempt to distill traces
of physics beyond the SM in the scenario that the
LHC only discovers a Higgs boson after its initial
years of operation with 10 $\invfb$ of data.
We focus on the scenario of light Higgs boson and used
the signal $\HGG$ to distinguish between
LHT, MUED, and the MSSM.
For simplicity, in the MSSM we consider a limited lone Higgs scenario
where the s-top soft masses are large enough that,
in addition to evade direct discovery at the LHC,
the s-top contributions to the gluon-gluon fusion and di-photon decay
amplitudes decouple regardless of the value of $A_t$.
Given the expected accuracies of the
measurement of the Higgs-di-gluon and Higgs-di-photon
couplings at this stage of operation of LHC,
and the implications of lone Higgs scenario
on the spectra of new physics,
in MUED it is difficult to have
significant deviations from the SM.
In the cases of
the LHT and the MSSM, however, it is possible to have a significant
suppression in the signal while discovering a lone
Higgs boson, and such lone Higgs scenario may give
rise to a new resonance
($H/A$ in the MSSM, and $W_{-}$ or $Q_{-}$ in the LHT)
with more luminosity.

In the case where the measured
deviation is very large,
we offer tests that may potentially separate
LHT from the MSSM even before a new resonance
is discovered.
In case where the deviation is small,
our work may nevertheless be useful and
we point out that the deviations
can be used to bias a class of model from
another,
and as such can be useful in
devising search strategies of the
favored new physics.

Although we have only worked out detailed strategies for
10 fb$^{-1}$ integrated luminosity in the case of lone Higgs scenario,
similar kinds of considerations could also apply to various stages of
LHC running.
We hope to have demonstrated that even with the lone Higgs scenario,
many insights can still be learned before the end of the
LHC era.

\begin{acknowledgments}
This work is supported by the US National
Science Foundation under grants
PHY-0354226, PHY-0555545, and PHY-0354838.
We thank Chuan-Ren Chen for carefully proofreading this preprint.
KH would like to thank R.~Sekhar Chivukula, Elizabeth H.~Simmons and
Neil D.~Christensen for stimulating discussions.
We would also like to thank Michael Spira
for his assistance with \verb"hdecay".
We thank Qing-Hong Cao, Debajyoti Choudhury, and
Sven Heinemeyer for granting us permission
to use the figures of their works.
\end{acknowledgments}

\end{document}